\documentclass[twocolumn]{aastex701}
\usepackage{graphicx}
\usepackage{subfigure}
\usepackage{amsmath}    
\usepackage{amssymb}    
\usepackage{bm}
\usepackage{url}
\usepackage{upgreek}
\usepackage{xspace}
\usepackage{mathtools}
\usepackage{booktabs}

\newcommand{\mdeg}{\ensuremath{^{\circ}}\xspace}

\newcommand{\ppcc}{$\,$pc$\,$cm$^{-3}$\xspace} 

\def\apsr{PSR~J1840+1102\xspace}
\def\bpsr{PSR~J1827$-$0849\xspace}
\def\cpsr{PSR~J1924+2027\xspace}

\graphicspath{{./}{figures/}{newfigs/}}

\begin{document}
\title{Survey of compact sources for pulsars and exotic objects - I. Overview and initial discoveries}
\author[orcid=0000-0002-0862-6062,sname='Maan']{Yogesh Maan}
\affiliation{National Centre for Radio Astrophysics (NCRA - TIFR), Pune - 411007, India}
\email[show]{ymaan@ncra.tifr.res.in}  

\author[orcid=0000-0002-2864-4110, sname='Bera']{Apurba Bera} 
\affiliation{ASTRON, the Netherlands Institute for Radio Astronomy, Oude Hoogeveensedijk 4,7991 PD Dwingeloo, The Netherlands}
\affiliation{International Centre for Radio Astronomy Research, Curtin University, Bentley, WA 6102, Australia}
\affiliation{Inter-University Centre for Astronomy and Astrophysics, Pune 411007, India}
\affiliation{National Centre for Radio Astrophysics (NCRA - TIFR), Pune - 411007, India}
\email{}

\author[orcid=0000-0001-5470-305X, sname='Lal']{Dharam Vir Lal}
\affiliation{National Centre for Radio Astrophysics (NCRA - TIFR), Pune - 411007, India}
\email{}

\author[0000-0002-5342-163X]{Yash Bhusare}
\affiliation{National Centre for Radio Astrophysics (NCRA - TIFR), Pune - 411007, India}
\email{}

\author[0000-0003-3203-1613]{Preeti Kharb}
\affiliation{National Centre for Radio Astrophysics (NCRA - TIFR), Pune - 411007, India}
\email{}

\author[0009-0000-6605-3162]{Banshi Lal}
\affiliation{National Centre for Radio Astrophysics (NCRA - TIFR), Pune - 411007, India}
\email{}

\author[0000-0001-8125-5619]{Pikky Atri}
\affiliation{ASTRON, the Netherlands Institute for Radio Astronomy, Oude Hoogeveensedijk 4,7991 PD Dwingeloo, The Netherlands}
\affiliation{Department of Astrophysics/IMAPP, Radboud University, P.O. Box 9010, 6500 GL, Nijmegen, The Netherlands}
\email{}
\correspondingauthor{Yogesh Maan}

\begin{abstract}
Targeted searches for pulsars based on their counterparts in radio images have resulted in the discovery of interesting pulsars including the first ever discovered millisecond pulsar (MSP). Here, we report the first results from our image-based survey of compact sources for pulsars and exotic objects (SCOPE). SCOPE utilizes interferometric as well as time-domain observations to search for radio pulsations as well as characterize the sources in the image-domain to identify their true nature. In the first stage of the SCOPE survey, we have used the Giant Metrewave Radio Telescope (GMRT) and the Green Bank Telescope (GBT) to follow up a sample of 31 compact and steep-spectrum sources. We provide an overview of the survey, the sample selection, the search procedures, and present discoveries of two MSPs --- \apsr and \bpsr. \apsr is a 1.6\,ms pulsar at the edge of the Scutum-Centaurus arm, while \bpsr is the radio counterpart of a gamma-ray pulsar that was earlier thought to be radio-quiet, and both the sources have very steep radio spectra. Using the interferometric data, we also provide a morphological classification of all the sources, model and characterize their spectra and identify the resolved, extragalactic sources in our sample. We discuss these results in the context of future image-based pulsar surveys.
\end{abstract}

\section{Introduction}
Radio pulsars are fast rotating neutron stars with exotic properties. A major fraction of radio pulsars have been discovered by blind surveys. These blind surveys are often agnostic to the exact sky locations, except for broad preferences like surveying the Galactic Plane where the majority of the pulsars are found. Targeted surveys, on the other hand, have some a priori information on the sky positions where pulsars are likely to be found. Targeted surveys have resulted in the discovery of a significant number of pulsars. Searches for pulsars in the Globular clusters is an example of one of the most successful targeted pulsar surveys --- more than 350 pulsars\footnote{See \url{https://www3.mpifr-bonn.mpg.de/staff/pfreire/GCpsr.html} for an up to date list of Globular cluster pulsars and the references listed therein.} have been discovered in 46 Globular clusters so far! The Large Area Telescope onboard the Fermi satellite (Fermi-LAT), with its large sensitivity, has revolutionized the field of gamma-ray pulsars \citep{Smith23}. Fermi-LAT discovered gamma-ray pulsars and the unassociated gamma-ray sources have also provided useful targets for several targeted radio surveys at diverse radio frequencies \citep[see, e.g.,][]{Ransom2011,Kerr2012,MA14}. Identifying the optical counterparts, primarily for the black widow and the spider millisecond pulsars (MSPs), and the information regarding their orbits therefrom, has also helped in uncovering radio pulsations from several interesting binary MSPs.
\par
Radio pulsars are one of the few known classes of astronomical sources that exhibit significant linear and circular polarization, steep spectra and are compact. These properties have been the basis for identification of pulsars in radio images. In fact, early discoveries of several interesting pulsars, including the first ever discovered MSP \citep[PSR~B1937+21;][]{Backer82} and the first pulsar discovered in a Globular cluster \citep[PSR~B1821$-$24;][]{Lyne87}, were made following their identification in the radio images as steep spectrum sources. \citet{Bhakta17} and \citet{Frail18} are some of the more recent, successful efforts where 8 pulsars have been discovered from the steep-spectrum sources first identified from the radio images. These discoveries demonstrate the potential of image-based targeted pulsar surveys.
\par
There have been several other image-based pulsar searches which did not result in discovery of any new pulsars \citep[e.g.,][and many others]{Maan18,Hyman19,Crawford21,Crawford25}. Most of those efforts were predominantly limited by the relatively coarse angular resolutions of the imaging surveys that were used to identify the potential pulsar candidates --- TIFR GMRT Sky Survey (TGSS) at 147\,MHz \citep[][]{Sirothia14,deGasperin18} and the NRAO VLA Sky Survey (NVSS) at 1.4\,GHz \citep{Condon98}, with typical angular resolutions of 25$\arcsec$ and 45$\arcsec$, respectively. The coarse angular resolutions resulted in a large contamination of the pulsar candidates by the background, extragalactic sources. More recently, images from the Rapid ASKAP Continuum Survey (RACS) have also become public. More specifically, RACS probes the radio sky at intermediate frequencies of 887\,MHz and 1367\,MHz \citep[RACS-low and RACS-mid, respectively;][]{racs1,racs4}, the latter is comparable to the frequency at which NVSS was conducted. More importantly, RACS images provide a much better angular resolution (around 8$\arcsec$ at 887\,MHz) and thus enable identification of some of the background extragalactic sources.
\par
Here, we present an image-based survey, survey of compact sources for pulsars and exotic objects (SCOPE), that employs a multi-domain (time as well as image domain) look at each candidate source and includes pulsar searches at multiple, suitable frequencies. The radio pulsation searches at multiple frequencies help in detecting any underlying pulsars in the presence of effects like intrinsic or propagation based pulse broadening. In the first stage of the SCOPE survey, we have examined a sample of compact and steep spectrum sources obtained utilizing the TGSS, NVSS and RACS surveys (more details in the following Section). The high-resolution image domain investigation from the SCOPE survey also identifies compact, steep spectrum sources that are not found to be pulsating. Apart from pulsars, high-redshift radio galaxies (HzRGs) are also often identified in the radio images as compact sources with steep spectra. Particularly at redshifts $>$ 6 (the epoch of re-ionization), HzRGs are of great interest as they provide tools to study the re-ionization of the Universe by active galaxies (e.g., accreting super-massive black holes). While pulsars could be confirmed using radio pulsation searches, confirming a HzRG would require high-resolution radio imaging, precise localization and actual measurements of the redshift using suitable optical spectroscopy or photometry measurements. Moreover, the high-resolution images also help in identifying and studying other interesting extragalactic objects.
\par
The rest of the paper is structured as follows. Details of the sample selection and observations are provided in Section 2, followed by analysis details in Section 3. Results from imaging as well as pulsar searches are provided in Section 4, including detailed accounts of individual pulsar discoveries, followed by a discussion and summary in Section 5.


\section{Sample selection and observations} \label{sec-obs}
\subsection{Sample selection}

The sample of compact and steep spectrum sources used in this work resulted from the initial stages of the work presented in \citet{SMB26}. Briefly, that work analyzed the publicly available images from the RACS survey at the sky positions of 171 sources that were found to be compact at 150\,MHz in the TGSS survey but not detected in the NVSS survey, and thus had spectral indices, $\alpha, <-1.5$ and $<-1.8$ in the Galactic Plane and the off-Galactic Plane, respectively. The spectral index limits were used directly from the catalogue by \citet[][]{deGasperin18}. From this analysis, 33 sources were identified in the initial stages, which were not detected in the RACS-low images at 887\,MHz, and the upper limits obtained from the RACS-low images and their TGSS flux densities indicated their spectra to be steep with indices $\alpha$$<$$-1.8$. Later, with the availability of more images from the RACS-mid survey and with improved analysis, some of the above 33 sources could be detected. The compactness and spectral indices were obtained for the sources that were detected in RACS. Overall, even for the sources which were not detected in RACS, better upper limits on their flux densities and hence, their spectral indices could be obtained. Two of the sources, J1751$-$2735 and J1901$-$0125 were recently identified as pulsars \citep{Bhakta17,Frail18} and these were dropped from the sample. All the remaining 31 sources that were identified in the initial stage, constitute the sample used in this work (see Table~\ref{tab:obs_summary}). These sources are distributed between declinations $-40\mdeg$ and $+40\mdeg$, primarily limited by the sky coverage of RACS. All these sources were observed, imaged and searched for radio pulsations.


\begin{table*}
\centering
\tabletypesize{\scriptsize}
\caption{Summary of the time-domain observations of all the sources. \label{tab:obs_summary}}
\begin{tabular}{lcccccccc}
\hline\hline
Source Name & RA & Dec & \multicolumn{2}{c}{300$-$500\,MHz (GMRT)} & \multicolumn{2}{c}{550$-$750\,MHz (GMRT)} & \multicolumn{2}{c}{720$-$920\,MHz (GBT)} \\
\cmidrule(lr){4-5}
\cmidrule(lr){6-7}
\cmidrule(lr){8-9}
 & (hh:mm:ss) & (dd:mm:ss) & N$_{\rm{chan}}$ & T$_{\rm{int}}$ (min) & N$_{\rm{chan}}$  & T$_{\rm{int}}$ (min) & N$_{\rm{chan}}$ & T$_{\rm{int}}$ (min) \\
\hline
S0112+0001 & 01:12:30 & +00:01:03 & 4096 & 10+7 & 2048 & 15+15 & \nodata & \nodata \\
S0404$-$1107 & 04:04:07 & $-$11:07:30 & 4096 & 15+15 & 2048 & 14+20 & \nodata & \nodata \\
S0412$-$0056 & 04:12:23 & $-$00:56:20 & 4096 & 10+10 & 2048 & 14+15 & 4096 & 27 \\
S0412$-$0101 & 04:12:23 & $-$01:01:46 & 4096 & 10+10 & 2048 & 15+15 & 4096 & 22 \\
S0501+1640 & 05:01:59 & +16:40:18 & 4096 & 15+15 & 2048 & 20+20 & \nodata & \nodata \\
S0636+0926 & 06:36:53 & +09:26:18 & 4096 & 10+10 & 2048 & 15+15 & \nodata & \nodata \\
S0637+0938 & 06:37:39 & +09:38:01 & 4096 & 10+10 & 2048 & 15+15 & \nodata & \nodata \\
S0705$-$0525 & 07:05:34 & $-$05:25:28 & 4096 & 10+10 & 2048 & 20+20 & \nodata & \nodata \\
S0708$-$0229 & 07:08:00 & $-$02:29:31 & 4096 & 15+15 & 2048 & 9+15 & \nodata & \nodata \\
S0907$-$1339 & 09:07:27 & $-$13:39:14 & 4096 & 15+15 & 2048 & 20 & 4096 & 28 \\
S0922$-$1426 & 09:22:10 & $-$14:26:28 & 4096 & 10+9 & 2048 & 20 & 4096 & 30 \\
S1106$-$2112 & 11:06:39 & $-$21:12:48 & 4096 & 10+9 & 2048 & 20 & 4096 & 30 \\
S1115$-$1651 & 11:15:26 & $-$16:51:43 & 4096 & 15+15 & 2048 & 30 & 4096 & 45 \\
S1134$-$1731 & 11:34:39 & $-$17:31:26 & 4096 & 9+10 & 2048 & 20 & 4096 & 30 \\
S1357$-$0846 & 13:57:29 & $-$08:46:53 & 4096 & 12+11 & \nodata & \nodata & \nodata & \nodata \\
S1455$-$1112 & 14:55:49 & $-$11:12:08 & 4096 & 11+10 & \nodata & \nodata & 4096 & 6 \\
S1713$-$3252 & 17:13:59 & $-$32:52:54 & 4096 & 15+15 & 2048 & 30+30 & \nodata & \nodata \\
S1727$-$1609 & 17:27:59 & $-$16:09:10 & 4096 & 5+5 & 2048 & 15+15 & \nodata & \nodata \\
S1733$-$3126 & 17:33:24 & $-$31:26:16 & 4096 & 15+15 & 2048 & 15+15 & \nodata & \nodata \\
S1747$-$3505 & 17:47:00 & $-$35:05:55 & 4096 & 15+15 & 2048 & 90+15+15 & 4096 & 101 \\
S1754$-$0841 & 17:54:19 & $-$08:41:35 & 4096 & 15+15 & 2048 & 15+15 & 4096 & 30 \\
S1759$-$0759 & 17:59:10 & $-$07:59:25 & 4096 & 15+15 & 2048 & 15+15 & 4096 & 30 \\
S1825+1246 & 18:25:06 & +12:46:16 & 4096 & 15+15 & 2048 & 15+15 & 4096 & 15 \\
S1827$-$0849 & 18:27:36 & $-$08:49:41 & 4096 & 15+15 & 2048 & 15+15+60 & 4096 & 90 \\
S1835$-$1421 & 18:35:45 & $-$14:21:39 & 4096 & 15+15 & 2048 & 15+15 & 4096 & 15 \\
S1839$-$1233 & 18:39:34 & $-$12:33:29 & 4096 & 15+15 & 2048 & 15+15 & 4096 & 15 \\
S1840+1102 & 18:40:09 & +11:02:07 & 2048 & 10+10 & 2048 & 15+30 & \nodata & \nodata \\
S1900+0859 & 19:00:42 & +08:59:19 & 4096 & 15+15 & 2048 & 15+15 & 4096 & 37 \\
S1924+2027 & 19:24:42 & +20:27:21 & \nodata & \nodata & 2048 & 15+15 & \nodata & \nodata \\
S2116$-$2053 & 21:16:54 & $-$20:53:19 & 4096 & 7+10 & 2048 & 15+15+50 & 4096 & 58 \\
S2222$-$0932 & 22:22:32 & $-$09:32:22 & 4096 & 8+10 & 2048 & 20+20 & \nodata & \nodata \\
\hline
\end{tabular}
\tablecomments{N$_{\rm{chan}}$ indicates the number of frequency channels. T$_{\rm{int}}$ shows the on-source integration time in minutes for each observation, with multiple observations separated by +.}
\end{table*}

\subsection{Observations}
The observations were conducted using the Giant Metrewave Radio Telescope (GMRT) and the Green Bank Telescope (GBT). GMRT consists of 30 dishes, each having 45\,m diameter, spread over nearly a 25\,km wide region. With GMRT, each source was observed twice, separately at each of the two frequency bands: band-3 (300$-$500\,MHz) and band-4 (550$-$750\,MHz). During each pointing, the interferometric as well as the phased-array (PA) beamformed filterbank data were recorded simultaneously. By coherently summing Nyquist-sampled voltages, the PA beam provides the sensitivity of a single aperture with a collecting area equal to the sum of all individual dishes, for the time-domain observations. In our observations, the PA beam data typically utilized 20$-$22 dishes (the central square dishes and a few dishes from each of the arms), while the interferometric data were recorded for all the available dishes. The interferometric data enable imaging of the field, while the PA data allow for radio pulsation searches. 
\par 
The interferometric data were recorded with typical sampling time of 10.7\,s at both the bands. Radio sources 3C286 (preferred whenever available), 3C48 and 3C147 were observed as the primary calibrator. A range of sources from the VLA calibrator list\footnote{\url{https://www.vla.nrao.edu/astro/calib/manual/csource.html}} were observed as secondary calibrators chosen based on proximity to the targets. The PA data were recorded with time resolutions of 81.92\,$\mu$s and 40.96\,$\mu$s at band-3 and band-4 respectively. The number of frequency channels were 4096 and 2048 at band-3 and band-4, respectively, and these were the same for the interferometric data as well. At most of the observing epochs, in addition to the PA beam, we also recorded the coherently dedispersed PA (CDPA) beam wherein the time-series data for the individual channels were coherently dedispersed at dispersion measures (DMs) of 75 and 225\,\ppcc at band-3 and band-4, respectively, before saving to the disk. The time resolution for these data were the same as in the original PA data, but the number of channels were reduced by a factor of two due to the backend's I/O budget limit. At some of the epochs, data from a post-correlation (PC) beam were recorded instead of the PA beam. The PC beam corresponds to phased-array data after subtracting the incoherent-array (IA) beam formed using the same dishes as the PA beam. For the typical number of dishes we used in our observations (20$-$22), the PA and PC beam theoretically provide similar sensitivities, however, the PC beam offers reduced baseline variations and some radio frequency interference (RFI) mitigation \citep{RCP18}.
\par
With a diameter of 100\,m, GBT is the largest steerable radio telescope. We used the 820\,MHz band of the GBT with 200\,MHz bandwidth for observations of a subset of the sample. The data were recorded in the search-mode filterbank format, with a time resolution of 81.92\,$\mu$s and 4096 channels across the 720$-$920\,MHz frequency range. Observation details of the filterbank data, including the observation durations and number of channels, for both the bands of GMRT as well as for the 820\,MHz band of GBT, are provided in Table~\ref{tab:obs_summary}.

\section{Analysis details} \label{sec-analysis}
\subsection{Imaging analysis}\label{sec-imag-analysis}
The interferometric data were processed using software packages \textsc{aips} \citep[Astronomical Image Processing System\footnote{\url{https://aips.nrao.edu/}};][]{aips90} and \textsc{casa} \citep[Common Astronomy Software Applications\footnote{\url{https://casa.nrao.edu/}};][]{casa2022} following standard imaging procedures \citep[see also][]{Lal2020,Lal2021}. After editing bad data, phases and amplitudes were calibrated using calibrator observations. The primary calibrators were also used as bandpass calibrators to correct the bandpass shape and set the flux density scale \citep{PerleyButler}. The bandpass calibrated visibilities were imaged using the task \textsc{tclean} in \textsc{casa}. We employed 3D imaging (gridder=`widefield'), two Taylor terms (nterms=2), and Briggs weighting (robust = 0 or 0.5) to ensure wide-field, wide-band image fidelity.
These initial images were then used to perform self-calibration of the visibilities, after which the final images were produced from the self-calibrated data, and corrected for the GMRT primary beam. The resultant angular resolution was $\approx 5 \arcsec$ in band-4 and $\approx 8 \arcsec$ in band-3; however the exact resolution varied from target to target. The uncertainty in the flux density scale, including calibration and systematic effects, is typically $\lesssim$5 per cent \citep{Lal2021}.


\subsection{Search for radio pulsations}
\subsubsection{Pre-processing}
The GMRT beamformed data (from either of the PA, PC or CDPA beams) recorded for both the bands are processed through a series of common steps. First, \texttt{RFIClean}\footnote{\url{https://github.com/ymaan4/RFIClean}} \citep{MvLV21} is used to excise periodic as well as narrow-band and spiky RFI, and then to convert the data to SIGPROC filterbank format. Then, \texttt{digifil} from \texttt{DSPSR} is used for down-sampling from 16\,bits to 8\,bits per sample. The effect of varying gain of the band-pass filter across the recorded bandwidth is also largely corrected by \texttt{digifil}. The resulting data are then subjected to \texttt{rfifind} from the pulsar exploration and search toolkit \citep[\textsl{PRESTO};][]{Ransom02} to generate RFI masks which are then used to exclude the RFI-contaminated time and frequency sections during any further processing.
\par
The GBT data were recorded in search-mode full-polarization PSRFITS format, and \texttt{psrfits\_subband} was used to reduce the data to total intensity, followed by \texttt{digifil} (from \textsl{DSPSR}) to write out SIGPROC filterbank output. Then, \texttt{rfifind} is used on the filterbank files to generate RFI masks.
\par
\begin{deluxetable*}{ccccccccc}
\tablecaption{The dedispersion plan used for the three observing bands. \label{tab:dm_search}}
\tablewidth{0pt}
\tablehead{
\multicolumn{3}{c}{300$-$500\,MHz (GMRT)} & \multicolumn{3}{c}{550$-$750\,MHz (GMRT)} & \multicolumn{3}{c}{720$-$920\,MHz (GBT)} \\
\cmidrule(lr){1-3} \cmidrule(lr){4-6} \cmidrule(lr){7-9}
\colhead{DM range} & \colhead{DM step} & \colhead{DS} & \colhead{DM range} & \colhead{DM step} & \colhead{DS} & \colhead{DM range} & \colhead{DM step} & \colhead{DS} \\
\colhead{(\ppcc)} & \colhead{(\ppcc)} & \colhead{factor} & \colhead{(\ppcc)} & \colhead{(\ppcc)} & \colhead{factor} & \colhead{(\ppcc)} & \colhead{(\ppcc)} & \colhead{factor}
}
\startdata
0$-$50   & 0.01 & 1  & 0$-$51    & 0.01 & 1  & 0$-$418   & 0.05 & 1 \\
50$-$100*  & 0.01 & 1  & 51$-$99   & 0.02 & 2  & 418$-$700 & 0.1  & 2 \\
100$-$150* & 0.03 & 2  & 99$-$150  & 0.05 & 4  & 700$-$1650 & 0.2 & 4 \\
150$-$395 & 0.05 & 8  & 150$-$198* & 0.03 & 2  &           &         &         \\
395$-$700 & 0.1  & 16 & 198$-$249* & 0.01 & 1  &           &         &         \\
          &         &         & 249$-$298* & 0.03 & 2  &           &         &         \\
          &         &         & 298$-$350* & 0.05 & 4  &           &         &         \\
          &         &         & 350$-$758 & 0.2  & 16 &           &         &         \\
          &         &         & 758$-$1350 & 0.5 & 32 &           &         &         \\
\enddata
\tablecomments{DS factor refers to the downsampling factor. An astrisk (*) implies that coherently dedispersed data (for DMs of 75\,\ppcc and 225\,\ppcc at bands-3 and 4, respectively) were used in these DM ranges.}
\end{deluxetable*}

\subsubsection{Search processing}
Each preprocessed data file is searched for periodic signals and bright radio bursts across a suitable range of DMs. At GMRT band-3, the PA beam data were used to prepare the dedispersed time-series in the DM ranges 0$-$50\,\ppcc and 150$-$700\,\ppcc, while the CDPA data were used in the DM range 50$-$150\,\ppcc, to minimize the dispersive smearing. Note that the band-3 CDPA beam was coherently dedispersed at 75\,\ppcc, and the number of channels in the CDPA beam was half that of the PA/PC beam. For band-3 data, the search was limited to a maximum DM of 750\,\ppcc as the expected scatter broadening towards most of the targets becomes very large (sometimes as large as a few seconds) beyond this DM. At GMRT band-4 and GBT 820\,MHz bands, the maximum DMs were adequately large. At GMRT band-4, typically the PA beam data were used to prepare the dedispersed time-series in the DM ranges 0$-$150\,\ppcc and 350$-$1350\,\ppcc, while the CDPA data were used in the DM range 150$-$350\,\ppcc, again to minimize the dispersive smearing (the band-4 CDPA beam was coherently dedispersed at a DM of 225\,\ppcc). The GBT 820\,MHz data were recorded only in the search mode (i.e., without any coherent dedispersion), and the search was conducted in the DM range 0$-$1650\,\ppcc. \texttt{DDplan.py} from \texttt{PRESTO} was used to make a dedispersion plan within the above specified individual DM ranges, separately for data at different bands. The dedispersion plans used for GMRT band-3 and band-4 data as well as the GBT 820\,MHz data are shown in Table~\ref{tab:dm_search}. The plan primarily includes the optimum DM step sizes used in different DM ranges, and the corresponding downsampling factors.
\par
The above dedispersion plan was used to prepare a number of incoherently dedispersed time-series using \texttt{prepsubband} from \texttt{PRESTO}. Each dedispersed time series is then searched for periodic signals using two rounds of \texttt{accelsearch} from \texttt{PRESTO}. In the first round, 16 harmonics are added and the acceleration parameter \texttt{zmax} is kept 0, primarily to search for periodic signals with narrow duty-cycles (often the case for normal, isolated pulsars). In the second round, \texttt{zmax} is fixed to 256 and 8 harmonics are added, which is more suitable to detect periodic signals affected by binary motion and with slightly larger duty cycles, often the case for MSPs. For each of the DM ranges shown in Table~\ref{tab:dm_search}, first the pulsar candidates were sorted according to their detection significance, and then 200 top candidates were folded using \texttt{prepfold} and the resulting diagnostic plots were examined by eye. Each observing session also involved a short scan on a known pulsar and the search pipeline was validated on these test pulsar scans. 
\par
Each of the dedispersed time series was also searched for any bright, sporadic pulses, using \texttt{single\_pulse\_search.py}. The resulting candidates are clustered in the time$-$DM space and the waterfall plots for the brightest candidates from each cluster were examined by eye. While the single pulse searches were performed towards each pointing, these searches in some of the GMRT observations were heavily affected by RFI.
\begin{figure*}[ht!]
\centering
\includegraphics[scale=0.5,trim={0.0cm 0 0.2cm 0},clip]{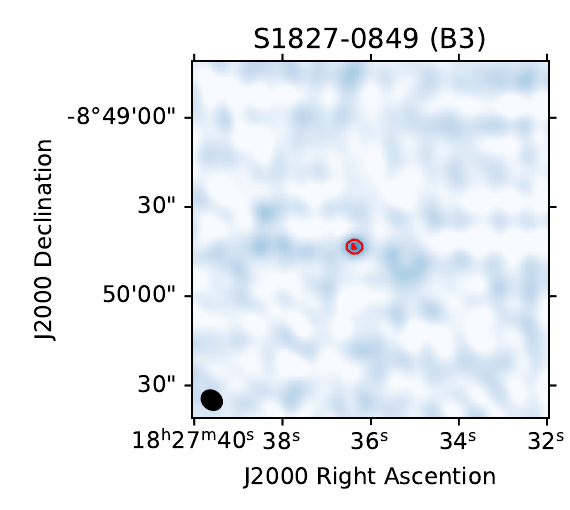}
\includegraphics[scale=0.5,trim={1.0cm 0 0.2cm 0},clip]{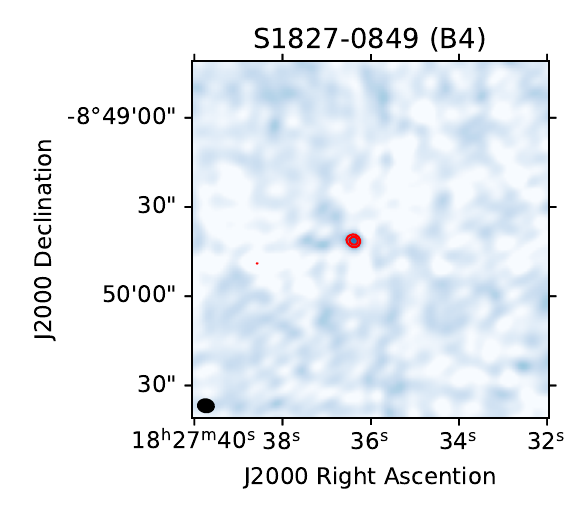}
\includegraphics[scale=0.5,trim={1.0cm 0 0.2cm 0},clip]{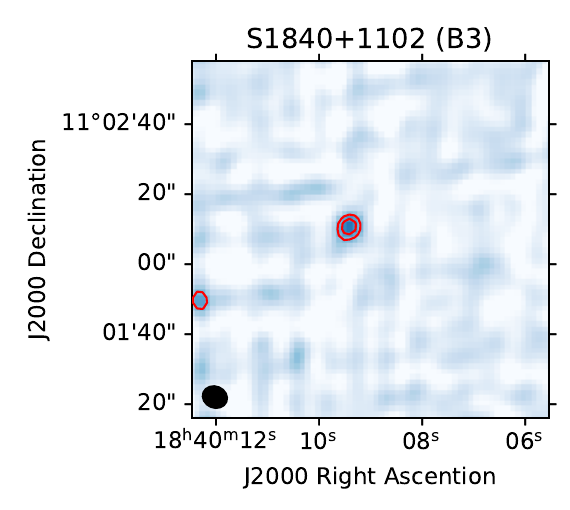}
\includegraphics[scale=0.5,trim={1.0cm 0 0.2cm 0},clip]{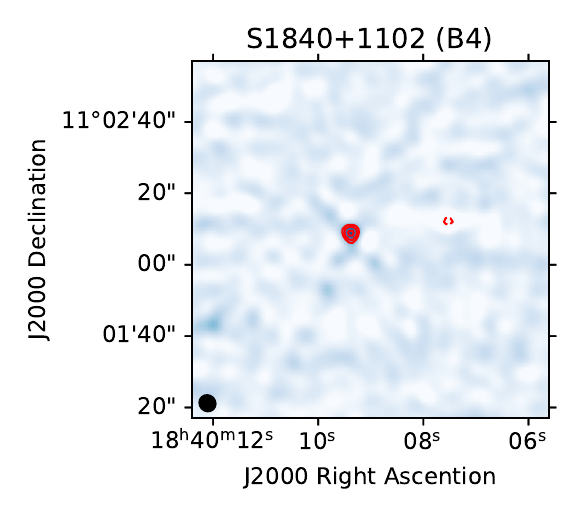}
\includegraphics[scale=0.5,trim={0.0cm 0 0.2cm 0},clip]{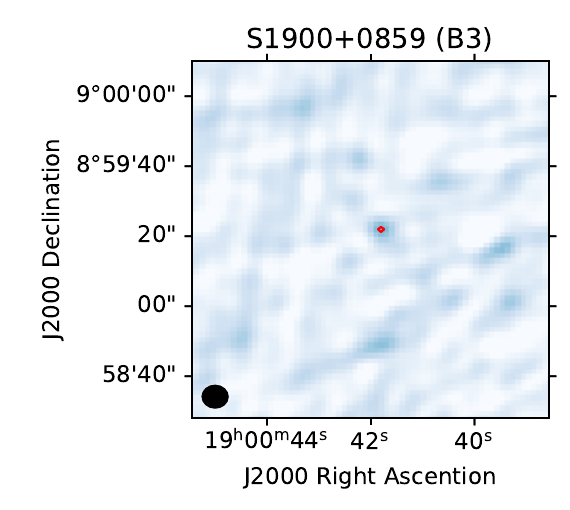}
\includegraphics[scale=0.5,trim={1.1cm 0 0.2cm 0},clip]{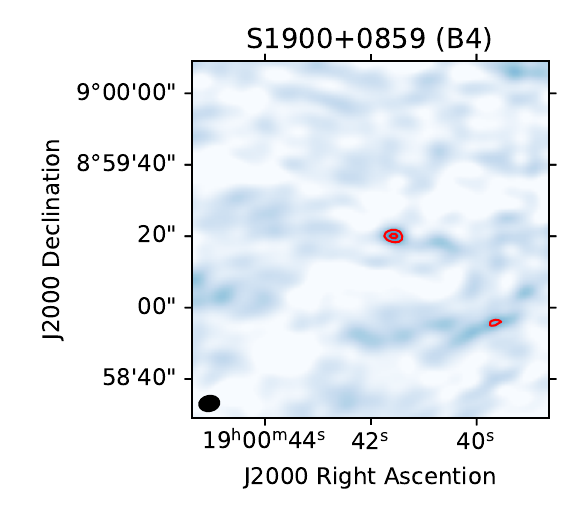}
\includegraphics[scale=0.5,trim={1.0cm 0 0.2cm 0},clip]{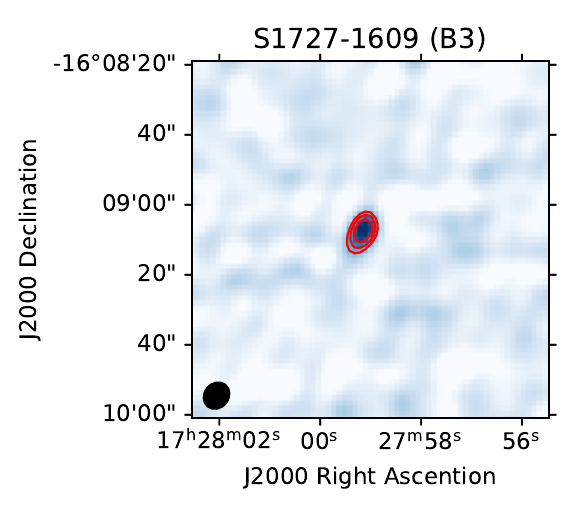}

\caption{Compact sources detected in band-3 and/or band-4. The color-scale is saturated at 10 times the local RMS noise. The lowest contour is at 4$\sigma_{local}$ level and the contour levels increase by factors of $\sqrt{2}$.
\label{fig-compact}}
\end{figure*}

\begin{figure*}[ht!]
\centering
\includegraphics[scale=0.5,trim={0.0cm 0 0.2cm 0},clip]{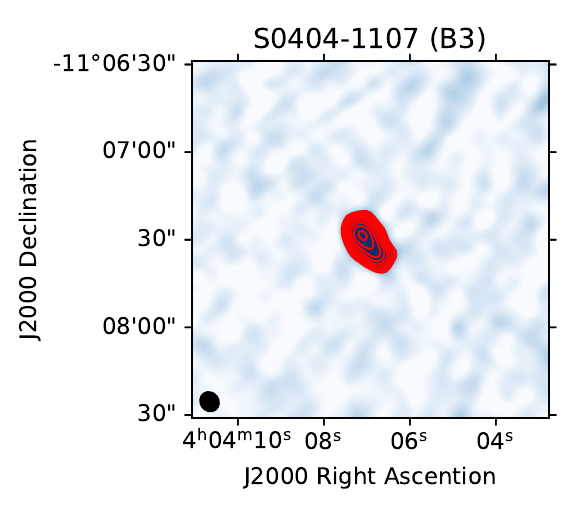}
\includegraphics[scale=0.5,trim={1.1cm 0 0.2cm 0},clip]{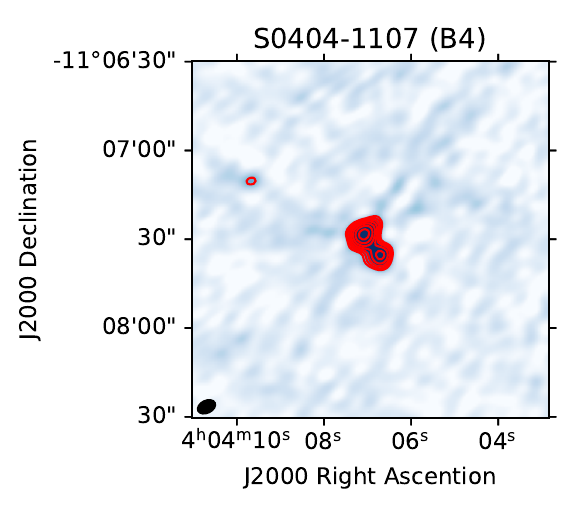}
\includegraphics[scale=0.5,trim={1.1cm 0 0.2cm 0},clip]{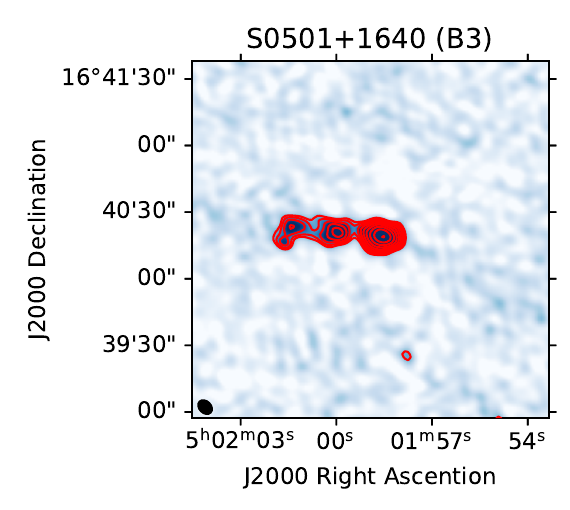}
\includegraphics[scale=0.5,trim={1.1cm 0 0.2cm 0},clip]{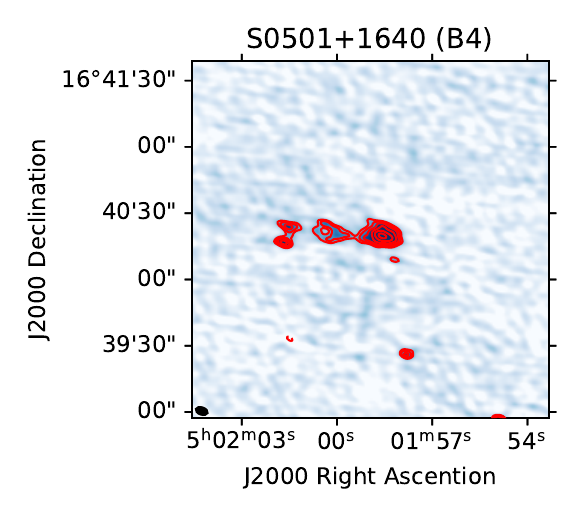}
\includegraphics[scale=0.5,trim={0.0cm 0 0.2cm 0},clip]{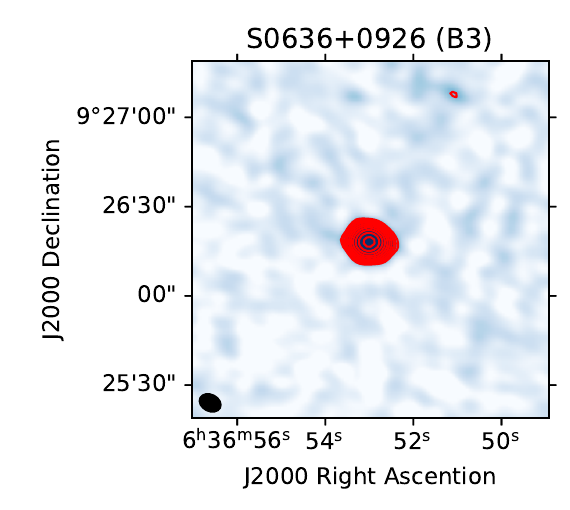}
\includegraphics[scale=0.5,trim={1.1cm 0 0.2cm 0},clip]{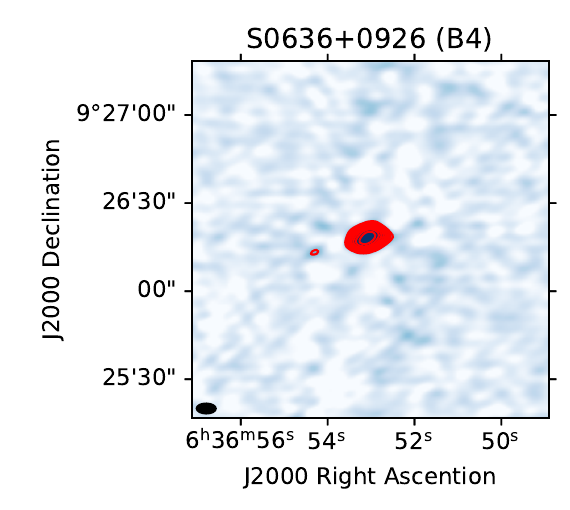}
\includegraphics[scale=0.5,trim={1.1cm 0 0.2cm 0},clip]{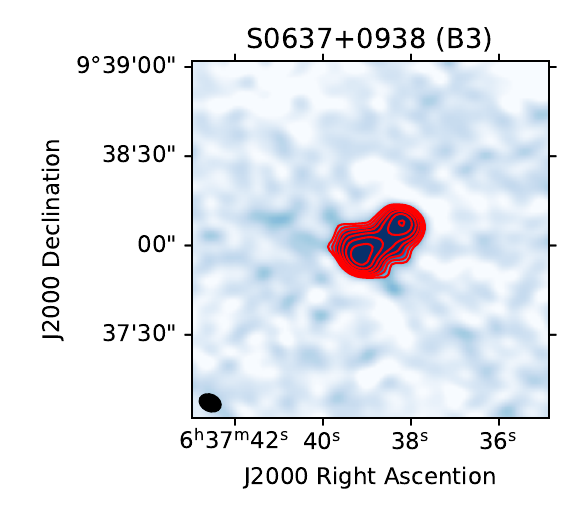}
\includegraphics[scale=0.5,trim={1.1cm 0 0.2cm 0},clip]{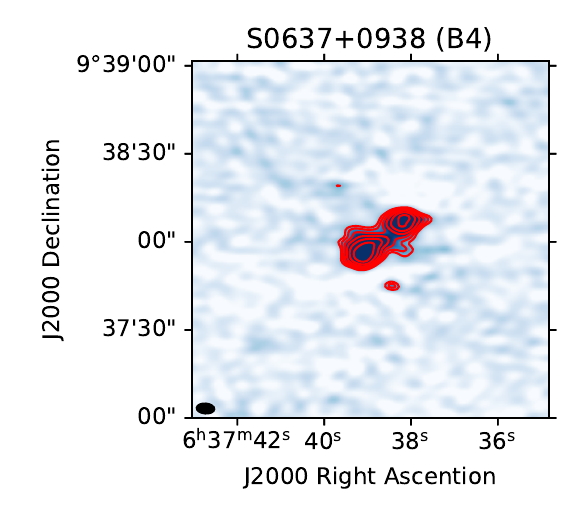}
\includegraphics[scale=0.5,trim={0.0cm 0 0.2cm 0},clip]{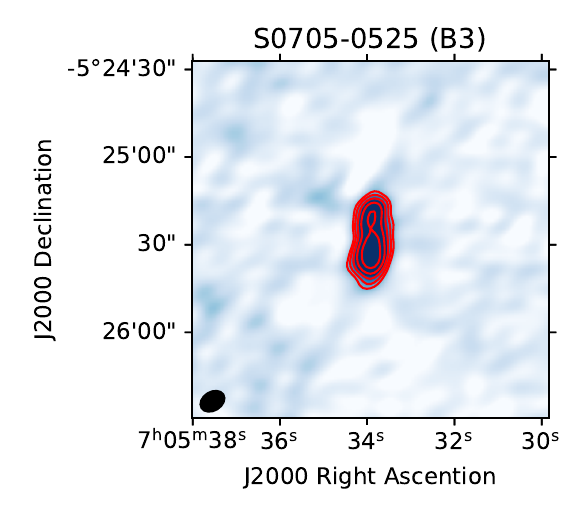}
\includegraphics[scale=0.5,trim={1.1cm 0 0.2cm 0},clip]{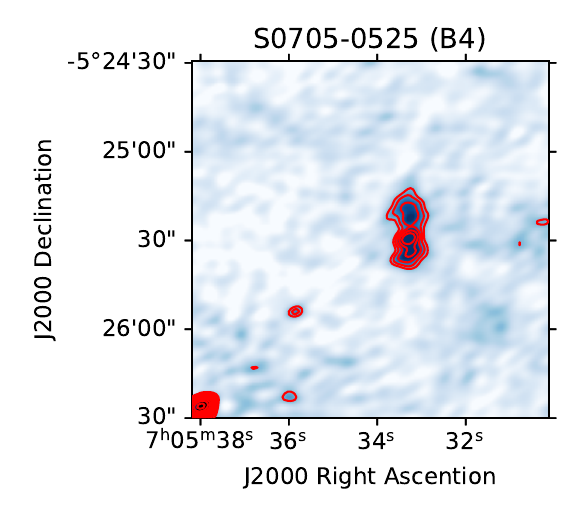}
\includegraphics[scale=0.5,trim={1.1cm 0 0.2cm 0},clip]{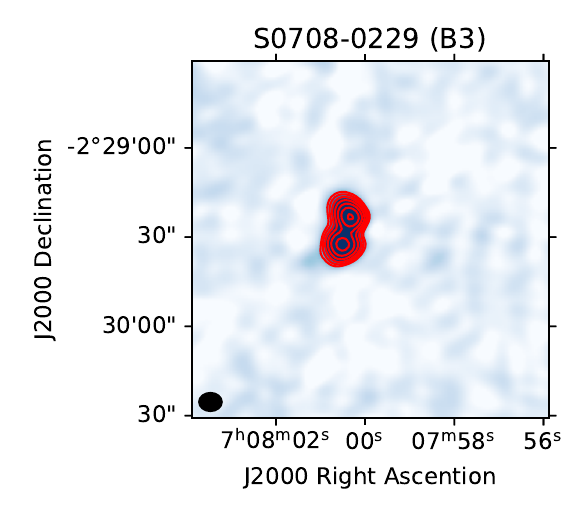}
\includegraphics[scale=0.5,trim={1.1cm 0 0.2cm 0},clip]{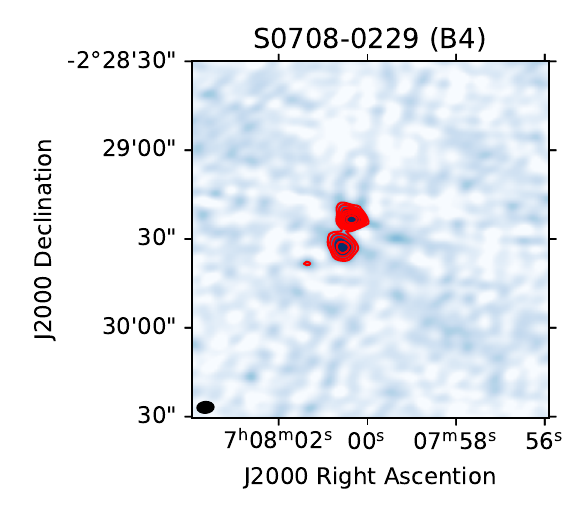}
\includegraphics[scale=0.5,trim={0.0cm 0 0.2cm 0},clip]{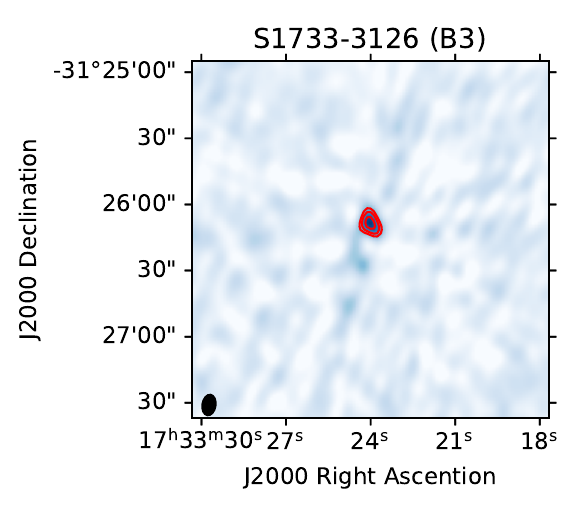} 
\includegraphics[scale=0.5,trim={1.1cm 0 0.2cm 0},clip]{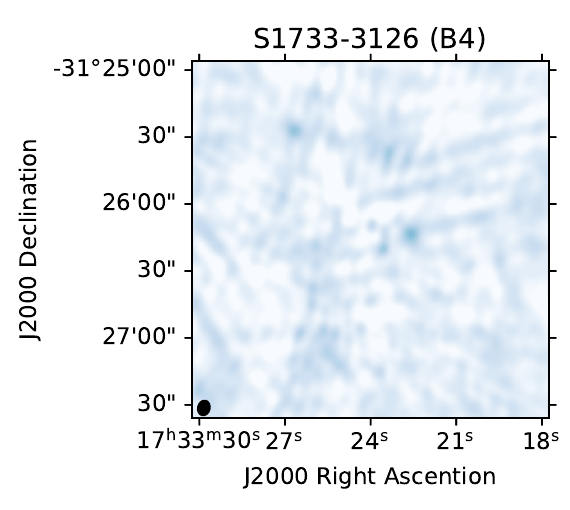}     
\includegraphics[scale=0.5,trim={1.1cm 0 0.2cm 0},clip]{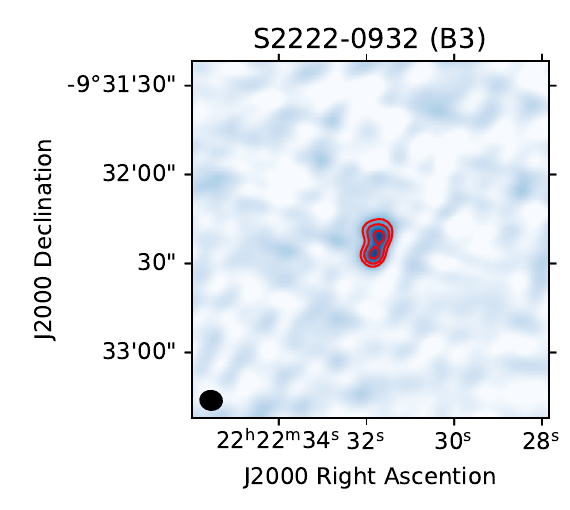}
\includegraphics[scale=0.5,trim={1.1cm 0 0.2cm 0},clip]{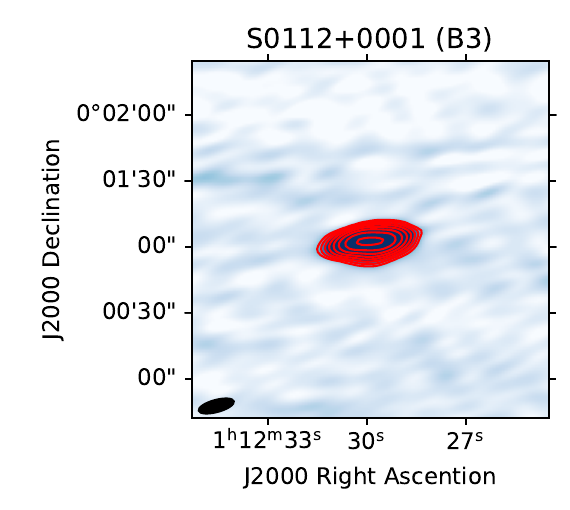}
\caption{Resolved sources detected in band-3 and/or band-4. The color-scale is saturated at 10 times the local RMS noise. The lowest contour represents 4$\sigma_{local}$ and the contour levels increase by factors of $\sqrt{2}$.
\label{fig-resolved}}
\end{figure*}

\section{Results} \label{sec-results}
\subsection{Imaging results}
Interferometric images of individual target fields, produced by the analysis described in Section~\ref{sec-imag-analysis}, were inspected for radio emission consistent with the location of the target sources in the TGSS catalogue. If any pixel within the TGSS synthesized beam (assumed to be circular with a diameter of 25\arcsec) was found to have an intensity larger than 4 times the local RMS noise ($\sigma_{\rm local}$), the target source is considered ``detected''. The local RMS noise ($\sigma_{\rm local}$) is estimated within a $ \approx 2\arcmin \times 2\arcmin$ region excluding visually identifiable sources, if any. For non-detected sources, classified as ``ND'', the detection threshold ($4\sigma_{\rm local}$) is quoted as an upper limit to the flux density assuming unresolved emission. For each detected target, the source was fit with an elliptical Gaussian using the \textsc{CASA} task \textsc{IMFIT} which returns estimates of the peak and the integrated flux densities. Sources with integrated flux densities larger than the peak by 3 times the standard deviation of the difference (estimated by adding the errors in quadrature), are classified as ``Resolved''. The remaining sources are classified as ``Compact''. The integrated flux densities are quoted as flux densities for ``Resolved'' sources, while the peak flux densities are quoted for the ``Compact'' ones.

 \par
For ``Resolved'' sources, the number of components are also noted, wherever possible, by visual identification. The images for the ``Compact'', ``Resolved'' and ``ND'' classes of the sources are shown in Figures~\ref{fig-compact}, \ref{fig-resolved} and \ref{fig-nondet}, respectively. For some of the sources, obtaining images with adequate quality was not possible due to excessive RFI contamination or due to the presence of a strong source in the close vicinity. The flux densities or the upper limits on flux densities at 400\,MHz and 650\,MHz obtained from the above imaging analysis, as well as those at 147\,MHz \citep[TGSS; from][]{deGasperin18}, 887\,MHz and 1367\,MHz (from RACS-low and RACS-mid), are listed in Table~\ref{tab:flux}. All the upper limits are quoted at four times the local rms, i.e., at $4\sigma_{local}$. For the individual sources, Table~\ref{tab:flux} indicates whether imaging was possible, separately at each of the two bands, in addition to any issues that might have compromised the image quality.  The table also lists the morphological classification of the sources. For the resolved sources, the letters ``S'', ``D'' and ``T'' indicate the presence of a single component, double components, and triple components, respectively. The table also lists the spectral indices or the limits obtained from the flux density measurements, more details of which are given in Section~\ref{sec-spidx}.
\begin{figure*}[ht!]
\centering
\subfigure{\includegraphics[width=0.25\textwidth]{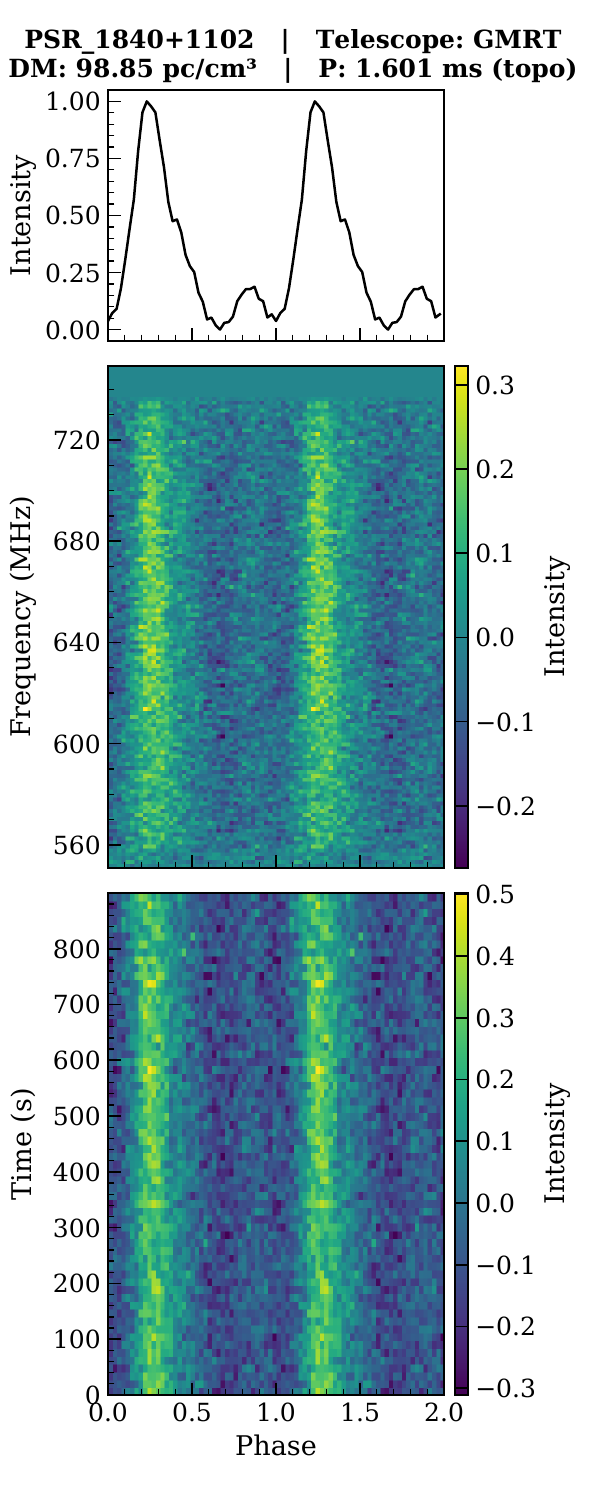} \label{fig-1840a}}
\subfigure{\includegraphics[width=0.25\textwidth]{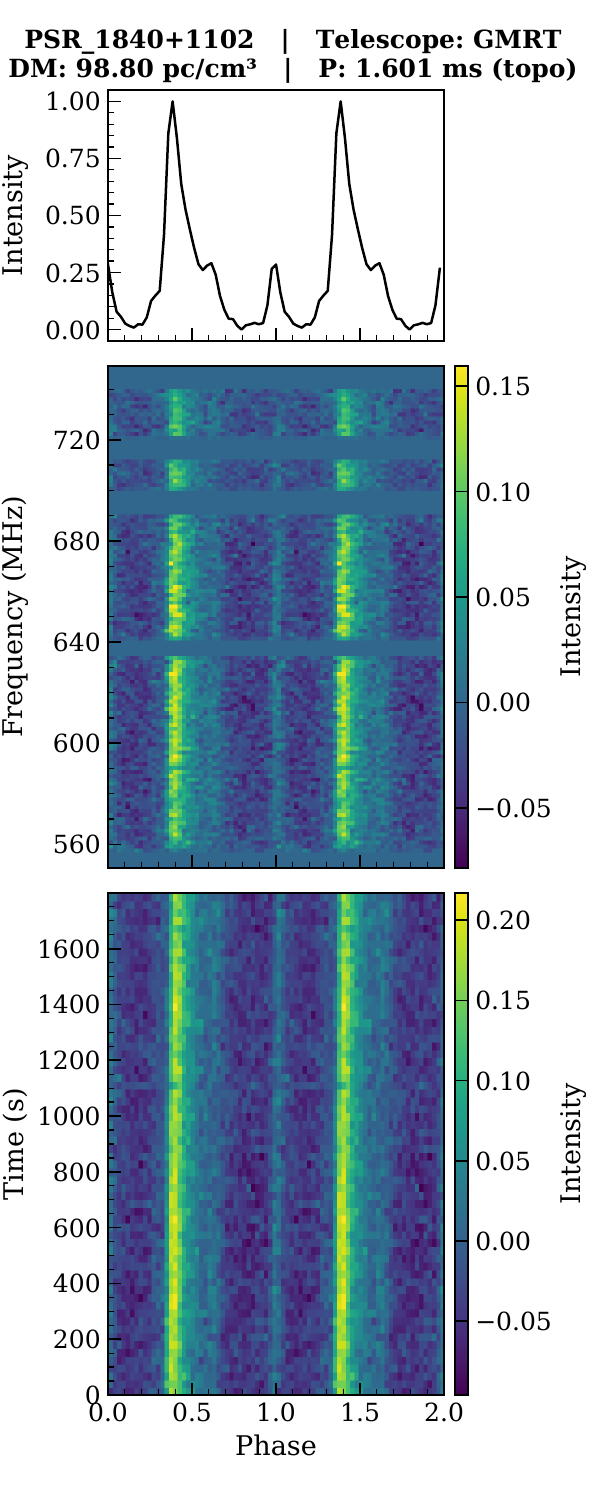} \label{fig-1840b}}
\subfigure{\includegraphics[width=0.25\textwidth]{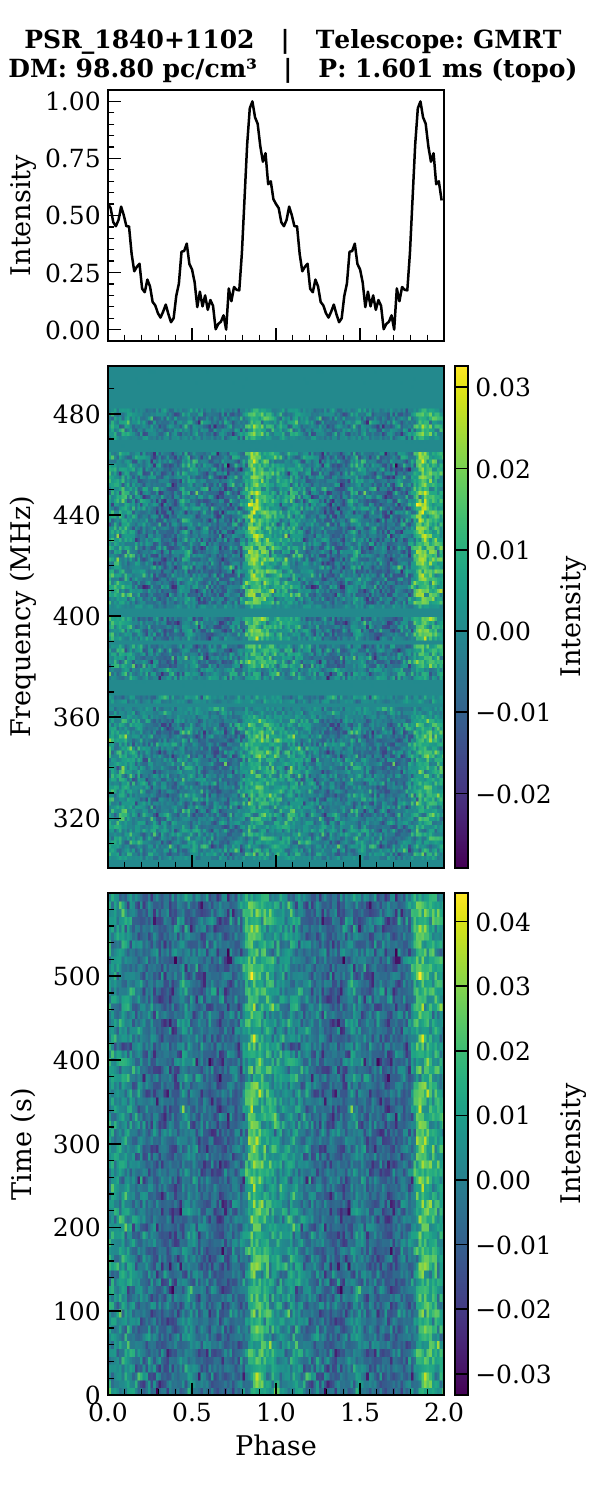} \label{fig-1840c}}
\caption{Summary diagnostic plots for the newly discovered pulsar \apsr at the discovery epoch (left; at GMRT band-4) and from the follow-up confirmations at band-4 (middle) and band-3 (right) of GMRT. Each summary plot shows the folded pulse intensity in the 2D spaces, rotation phase vs. time (bottom panel) and rotation phase vs. frequency (middle panel) spaces, and the average profile (top panel). The intensity in these plots is in arbitrary units and the peak is normalized to unity. \label{fig-1840}}
\end{figure*}


\begin{deluxetable*}{lccccccccc}
\tablecaption{Flux density measurements \label{tab:flux}}
\tablewidth{0pt}
\tablehead{
\colhead{Source} & \colhead{$S_{\rm 150}$} & \colhead{$S_{\rm 887}$} & \colhead{$S_{\rm 1367}$} & \colhead{$S_{\rm 400}$} & \colhead{$S_{\rm 650}$} & \colhead{B3} & \colhead{B4} & \colhead{Image} & \colhead{$\alpha$} \\
\colhead{} & \colhead{(mJy)} & \colhead{(mJy)} & \colhead{(mJy)} & \colhead{(mJy)} & \colhead{(mJy)} & \colhead{} & \colhead{} & \colhead{Class} & \colhead{}
}
\startdata
S0112+0001     & $98\pm6$   & $6.4\pm0.9$ & $1.4\pm0.3$ & $26.9\pm0.6$ & \nodata & Y & \nodata & Resolved (S) & $-1.5\pm0.2$ \\
S0404$-$1107   & $137\pm9$  & $3.6\pm0.8$ & $1.4\pm0.4$ & $16.4\pm0.4$ & $6.6\pm0.5$ & Y & Y & Resolved (D) & $-2.1\pm0.1$ \\
S0412$-$0056   & $200\pm16$ & $<2.6$      & $<0.9$      & $<10.4 $ & $<6.0 $ & Y* & Y* & ND & $\leq-3.0$ \\
S0412$-$0101   & $167\pm17$ & $<2.7$      & $<0.9$      & $<14.8 $ & $<7.6 $ & Y* & Y* & ND & $\leq-2.4$ \\
S0501+1640     & $94\pm10$  & $2.6\pm1.0$ & $<0.9$      & $9.7\pm0.7$ & $3.8\pm0.4$ & Y & Y & Resolved (T) & $-2.2\pm0.1$ \\
S0636+0926     & $52\pm7$   & $4.8\pm0.9$ & $2.2\pm0.6$ & $13.7\pm0.2$ & $7.3\pm0.2$ & Y & Y & Resolved (S) & $-1.3\pm0.1$ \\
S0637+0938     & $82\pm7$   & $8.9\pm0.9$ & $4.9\pm1.3$ & $19.1 \pm 1.5$ & $8.8\pm0.9$ & Y & Y & Resolved (D) & $-1.3\pm0.1$ \\
S0705$-$0525   & $53\pm7$   & $4.2\pm0.8$ & $1.0\pm0.3$ & $10.5\pm0.6$ & $5.4\pm0.4$ & Y & Y & Resolved (D) & $-1.5\pm0.1$ \\
S0708$-$0229   & $68\pm8$   & $4.7\pm0.7$ & $2.7\pm0.7$ & $9.1\pm0.6$ & $2.6\pm0.2$ & Y & Y & Resolved (D) & $-1.8\pm0.4$ \\
S0907$-$1339   & $145\pm26$ & $<1.9$      & $<0.6$      & \nodata & $<2.8$ & \nodata & Y* & ND & $\leq-2.5$ \\
S0922$-$1426 & $224\pm35$ & $<4.0$      & $<0.9$      & \nodata & $<24$ & \nodata & Y* & ND & $\leq-2.5$ \\
S1106$-$2112   & $223\pm17$ & $<3.4$      & $<1.4$      & \nodata & $<26$ & \nodata & Y* & ND & $\leq-2.3$ \\
S1115$-$1651   & $95\pm8$   & $<2.0$      & $<1.1$      & \nodata & $<5.2$ & \nodata & Y* & ND & $\leq-3.1$ \\
S1134$-$1731 & $167\pm14$ & $<2.1$      & $<0.8$      & \nodata & $<18$ & \nodata & Y* & ND & $\leq-2.4$ \\
S1357$-$0846 & $87\pm6$   & $1.8\pm0.6$ & $<0.8$      & \nodata & $<3.6$ & \nodata & Y* & Compact\dag & $-2.2\pm0.1$ \\
S1455$-$1112 & $134\pm17$ & $<5.4$      & $<1.5$      & \nodata & $<140$ & \nodata & Y* & ND & $\leq-2.0$ \\
S1713$-$3252   & $95\pm8$   & $9.5\pm1.1$ & $4.5\pm1.2$ & $<3.2 $ & $<0.9 $ & Y* & Y* & Resolved (D)\dag & $-2.0\pm1.0$ \\
S1727$-$1609   & $111\pm7$  & $1.3\pm0.4$ & $<1.6$      & $2.97\pm0.25$ & \nodata & Y & N & Compact & $-3.5\pm0.7$ \\
S1733$-$3126   & $114\pm12$ &$22.7\pm1.2$ & $11.4\pm0.8$ & $9.8\pm1.0$ & $<2.0 $ & Y* & Y* & Resolved (D)\dag & $-0.6\pm0.8$ \\
S1747$-$3505   & $84\pm12$  & $0.9\pm0.3$ & $<0.8$      & $<1.2 $ & $<0.6 $ & Y* & Y* & Compact & $-3.4\pm0.9$ \\
S1754$-$0841   & $91\pm13$  & $<1.7$      & $<0.7$      & $<3.8 $ & $<2.5 $ & Y* & Y* & ND & $\leq-3.2$ \\
S1759$-$0759   & $151\pm16$ & $<1.3$      & $<0.7$      & $<7 $ & $<3.5 $ & Y* & Y* & ND & $\leq-3.1$ \\
S1825+1246     & $98\pm12$  & $<1.7$      & $<0.9$      & $<38 $ & $<10.5 $ & Y* & Y* & ND & $\leq-2.3$ \\
S1827$-$0849   & $220\pm16$ & $<3.1$      & $<1.0$      & $2.4\pm0.4$ & $0.53\pm0.07$ & Y & Y & Compact & $-4.2\pm0.3$ \\
S1835$-$1421   & $54\pm7$   & $<2.0$      & $<1.2$      & $<28.0 $ & $<6.4 $ & Y* & Y* & ND & $\leq-1.8$ \\
S1839$-$1233   & $72\pm10$  & $<2.1$      & $<1.5$      & $<35 $ & $<4.8 $ & Y* & Y* & ND & $\leq-2.0$ \\
S1840+1102     & $100\pm12$ & $1.6\pm0.4$ & $<0.9$      & $ 3.0 \pm 0.4$ & $1.2 \pm 0.2^{**}$ & Y & Y & Compact & $-2.8\pm0.5$ \\
S1900+0859     & $130\pm17$ & $4.6\pm0.6$ & $1.4\pm0.3$ & $3.7\pm0.9$ & $1.3\pm0.2$ & Y & Y & Compact & $-2.1\pm0.6$ \\
S1924+2027     & $91\pm8$   & $<1.5$      & $<1.0$      & \nodata & $<1.4 $ & \nodata & Y & ND & $\leq-2.8$ \\
S2116$-$2053   & $92\pm12$  & $<2.5$      & $<1.3$      & $<3.4$ & \nodata & Y* & \nodata & ND & $\leq-3.3$ \\
S2222$-$0932   & $130\pm6$  & $<1.3$      & $<0.9$      & $9.9\pm1.3$ & \nodata & Y & \nodata & Resolved (D) & $-2.7\pm0.1$ \\
\enddata
\tablecomments{The five columns after the source name present the flux densities at 150~MHz (TGSS), 887~MHz and 1367~MHz (RACS), 400~MHz and 650~MHz (GMRT; this work). Values with ``$<$'' indicate upper limits, defined as 4 times the local RMS noise (i.e., $4\sigma$). The B3 and B4 columns indicate whether imaging of the source was possible at band-3 and band-4, respectively. An asterisk(*) in B3/B4 column denotes that the image quality near the target is compromised by the proximity of one or more bright sources, leading to a significantly lower sensitivity than other areas of the image. ** indicates the source to have variable flux density and the value provided here is estimated from a combined obtained from data over a few epochs. \dag~indicates that classification is based on RACS images.}
\end{deluxetable*}

\begin{deluxetable*}{ccccccccc}
\tablecaption{A summary of the parameters of pulsars discovered/re-detected in this work. \label{tab:psr_summary}}
\tablewidth{0pt}
\tablehead{
\colhead{PSR Name} & \colhead{RA} & \colhead{Dec} & \colhead{Period} & \colhead{DM} & Survey Component & \colhead{Remarks} \\
 & \colhead{(J2000)} & \colhead{(J2000)} & \colhead{(ms)} & \colhead{(\ppcc)} &  & 
}
\startdata
J1840+1102  & 18:40:09.6$\pm$0.9 & +11:02:10.6$\pm$0.6    & 1.60    & 98.8  &  GM-SCOPE  & Binary\\
J1827-0849  & 18:27:36.4$\pm$0.3 & $-$08:49:41.3$\pm$0.2  & 2.24    & 197.8 &  GB-SCOPE  & Gamma-ray pulsar \\
J1924+2027  & 19:24:42$\pm$0.1    & +20:27:21$\pm$2        & 1.95    & 211.7 &  GM-SCOPE  & A re-discovery; \citet{Han21} \\
\enddata
\tablecomments{J1924+2027 was not detected in the interferometric images, neither in those from this work nor in the RACS survey. Thus, its best position is obtained from TGSS, with assumed uncertainty of 2\arcsec.}
\end{deluxetable*}

\subsection{Pulsar search results}
As mentioned above, 9 sources turned out to be resolved at our angular resolutions of 8$\arcsec$ (band-3) or 4$\arcsec$ (band-4), and thus, are unlikely to be pulsars. The other 24 sources remain radio pulsar candidates. Our searches for radio pulsations from these sources resulted in the discovery of two new radio pulsars and a detection of another pulsar, which was discovered earlier by FAST. A summary of the parameters of these pulsars is provided in Table~\ref{tab:psr_summary}, and further details of these discoveries are provided below. To indicate the discovery telescope for individual pulsars, in Table~\ref{tab:psr_summary} we indicate the GMRT component of the survey by GM-SCOPE and GBT component by GB-SCOPE.
\begin{figure*}[ht!]
\centering
\subfigure[]{\includegraphics[width=0.25\textwidth]{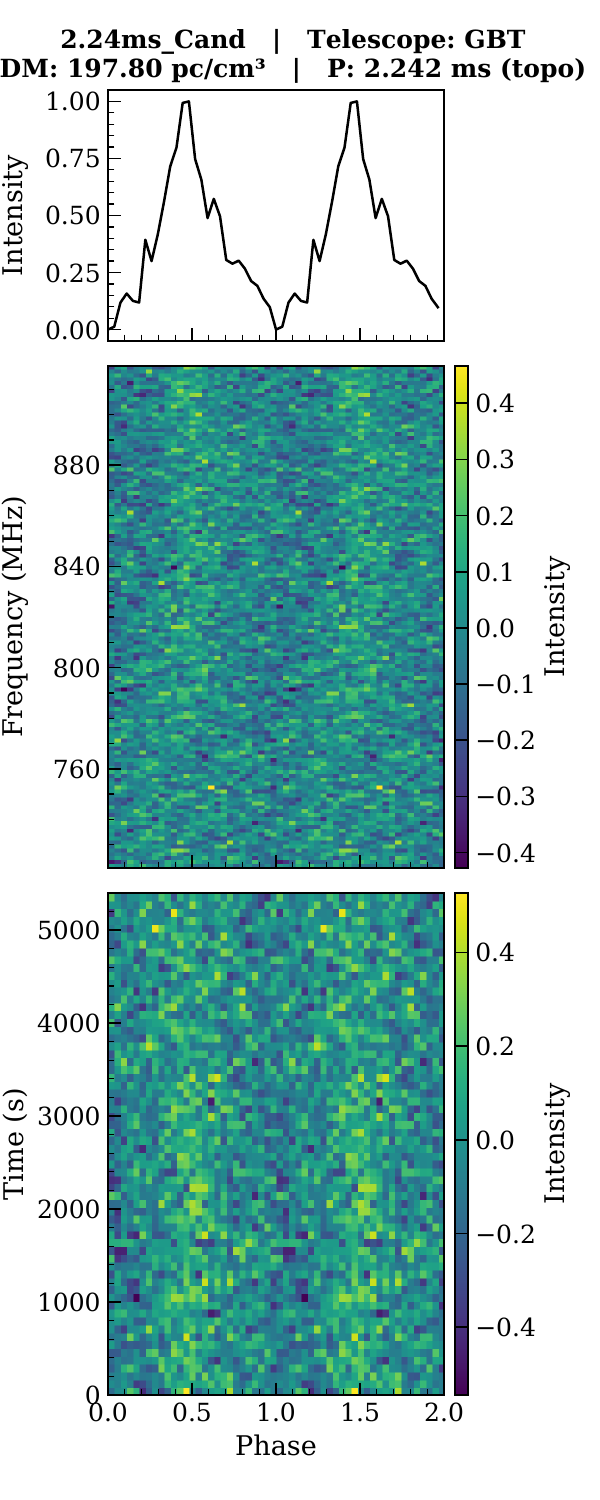}
\label{fig-1827a}}
\subfigure[]{\includegraphics[width=0.25\textwidth]{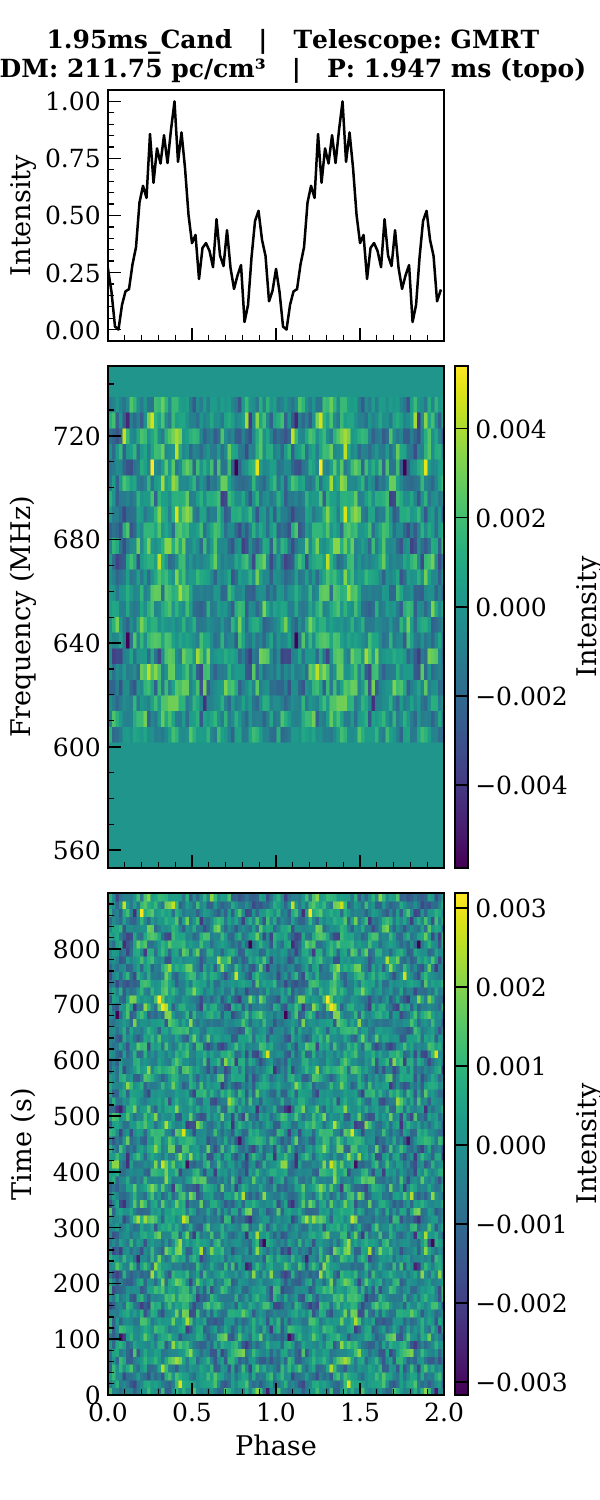}
\label{fig-1924a}}
\caption{Summary diagnostic plots for the newly discovered pulsar \bpsr at the discovery epoch (left; GBT 820\,MHz band) and the rediscovery of \cpsr (right; GMRT band-4). For the description of individual panels, please see Figure~\ref{fig-1840} caption. \label{fig-1827}}
\end{figure*}
\begin{figure*}[ht!]
\centering
\includegraphics[width=0.95\textwidth]{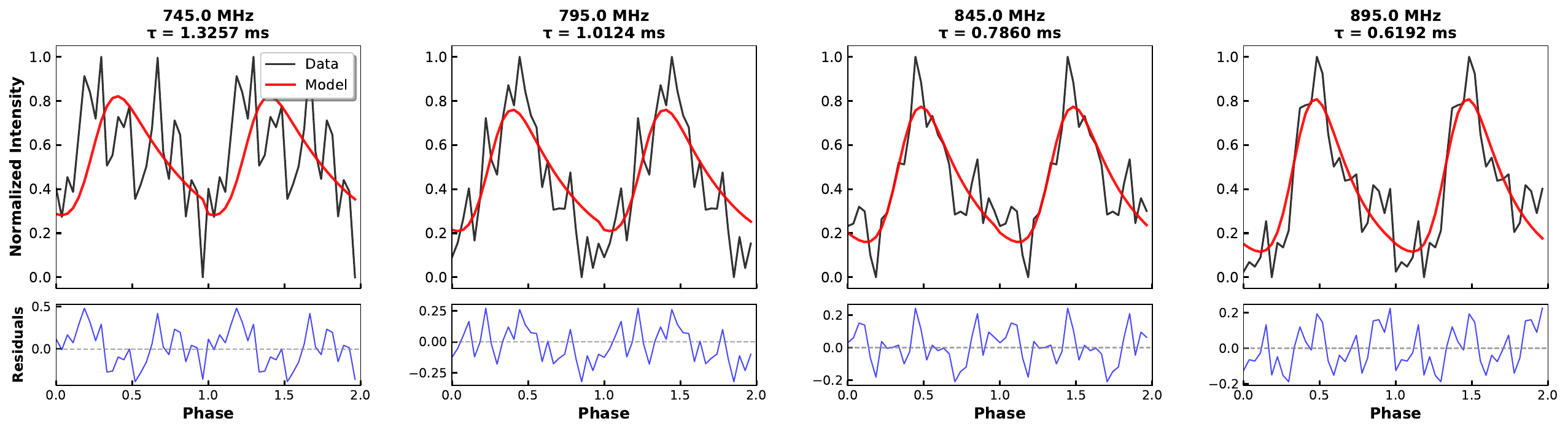}
\caption{Sub-banded profiles of \bpsr are shown in the upper panels (black) at the center frequencies indicated in the title of each panel, along with the modeled profiles in red. The residuals are shown in the bottom panels. \label{fig-1827scat}}
\end{figure*}
\subsubsection{\apsr}
This pulsar was discovered from an observation conducted on 09 May 2023 and the data were recorded only from the PC~beam. \apsr has a spin period of 1.6\,ms and it is the sixth fastest rotator among the known pulsars \citep{atnf}. The pulsar was discovered at a DM of 98.85\,\ppcc, which was refined to 98.80\,\ppcc in subsequent high time-resolution follow-up observations conducted using the CDPA beam at band-4 as well as band-3. Figure~\ref{fig-1840} shows the summary plots (pulse intensity in the two-dimensional spaces, phase$-$time and phase$-$frequency spaces, as well as the average profile) for \apsr at the discovery as well as the follow-up epochs. The profile appears more smooth and wider at the discovery epoch (left set of panels in Figure~\ref{fig-1840}) due to the dispersion smearing across the individual channels, which is estimated to be nearly 20\% of the spin period at 650\,MHz. The coherently dedispersed profile obtained from a follow-up observation at the same frequency (middle set of panels in Figure~\ref{fig-1840}) more clearly brings out the intrinsically narrow components in the profile. The radio emission covers a major fraction of the rotation phases. At band-3, an exponential scattering tail is also clearly visible, which causes the radio emission to sample the entire range of rotation phases. 

\subsubsection{\bpsr}
This source was identified as a steep-spectrum, Fermi unassociated LAT source and a likely pulsar \citep{Frail18}. Facilitated by the sky position from the radio images, a gamma-ray MSP was indeed uncovered \citep{Frail18}. However, the radio counterpart was not identified and it was included as a radio quiet pulsar in the third Fermi LAT catalog of gamma-ray pulsars \citep{Smith23}. Our searches in the 300-500\,MHz and 550-750\,MHz frequency ranges using GMRT also could not find any radio pulsations, despite having enough sensitivity. Using GBT at 820\,MHz, a deep integration revealed the radio counterpart, a 2.24\,ms pulsar at a DM of 197.8\,\ppcc. While the detected periodic signal is faint, with a S/N of around 10, it is consistent with the expected flux density at 820\,MHz. Moreover, the period confirms that it is indeed the radio counterpart of the gamma-ray pulsar. Figure~\ref{fig-1827} shows the diagnostic summary plot for \bpsr at the discovery epoch.
\par
The profile from our faint detection in the 720$-$920\,MHz band (see Figure~\ref{fig-1827}) exhibits a scattering tail. To estimate the scatter broadening, we modeled the profile in 4 different subbands, each with a single gaussian component but having a common width across subbands, and convolved with a one-sided exponential function whose characteristic width, $\tau$, changes with frequency as a power-law ($\tau \propto \nu^{\alpha}$). This modeling using \texttt{Scipy}'s \texttt{curve\_fit} results in $\tau=0.91\pm0.05$\,ms at a reference frequency of 816.5\,MHz and the power-law index, $\alpha=-4.2\pm0.3$. The modeled and the original profiles at different sub-bands are shown in Figure~\ref{fig-1827scat}. At the lowest sub-band centered at 745\,MHz, $\tau=1.3\pm0.2$\,ms and the scattering tail already extends beyond the rotation period.
\subsubsection{\cpsr}
\cpsr was already discovered by the FAST Galactic Plane pulsar snapshot (FAST-GPPS) survey \citep{Han21}. The source remained in the original list of our candidates as the sifting for known pulsars was done using an older version of the ATNF catalog. We re-discovered this pulsar using our observations conducted on 2023 May 05 (see Figure~\ref{fig-1827}). The S/N-maximizing period and DM are 1.95\,ms and 211.7\,\ppcc, respectively, and confirm it to be a low frequency detection of \cpsr. Given its relatively high DM value and faint periodic signal, this source was used as a test pulsar in some of our follow-up observations. The more precise sky position that we list in Table~\ref{tab:psr_summary} would help in obtaining a timing model of this pulsar.

\begin{figure*}[ht!]
\centering
\includegraphics[width=0.95\textwidth]{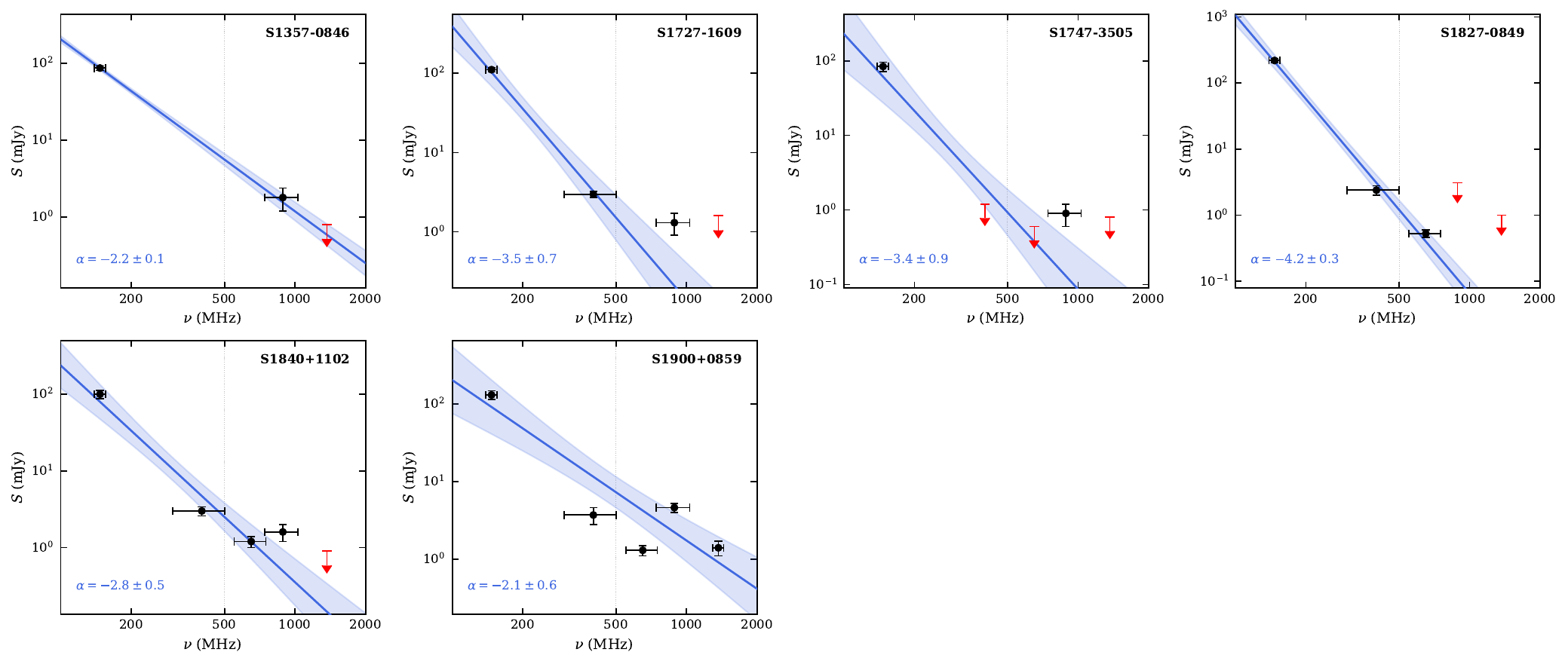}
\caption{The power-law spectral models of all the compact sources. The black and red points represent the measured flux densities and upper limits, respectively. The best fit model is shown as a blue line, and the uncertainty region is shown by the light blue shaded region. \label{fig-spidx-compact}}
\end{figure*}
\begin{figure*}[ht!]
\centering
\includegraphics[width=0.95\textwidth]{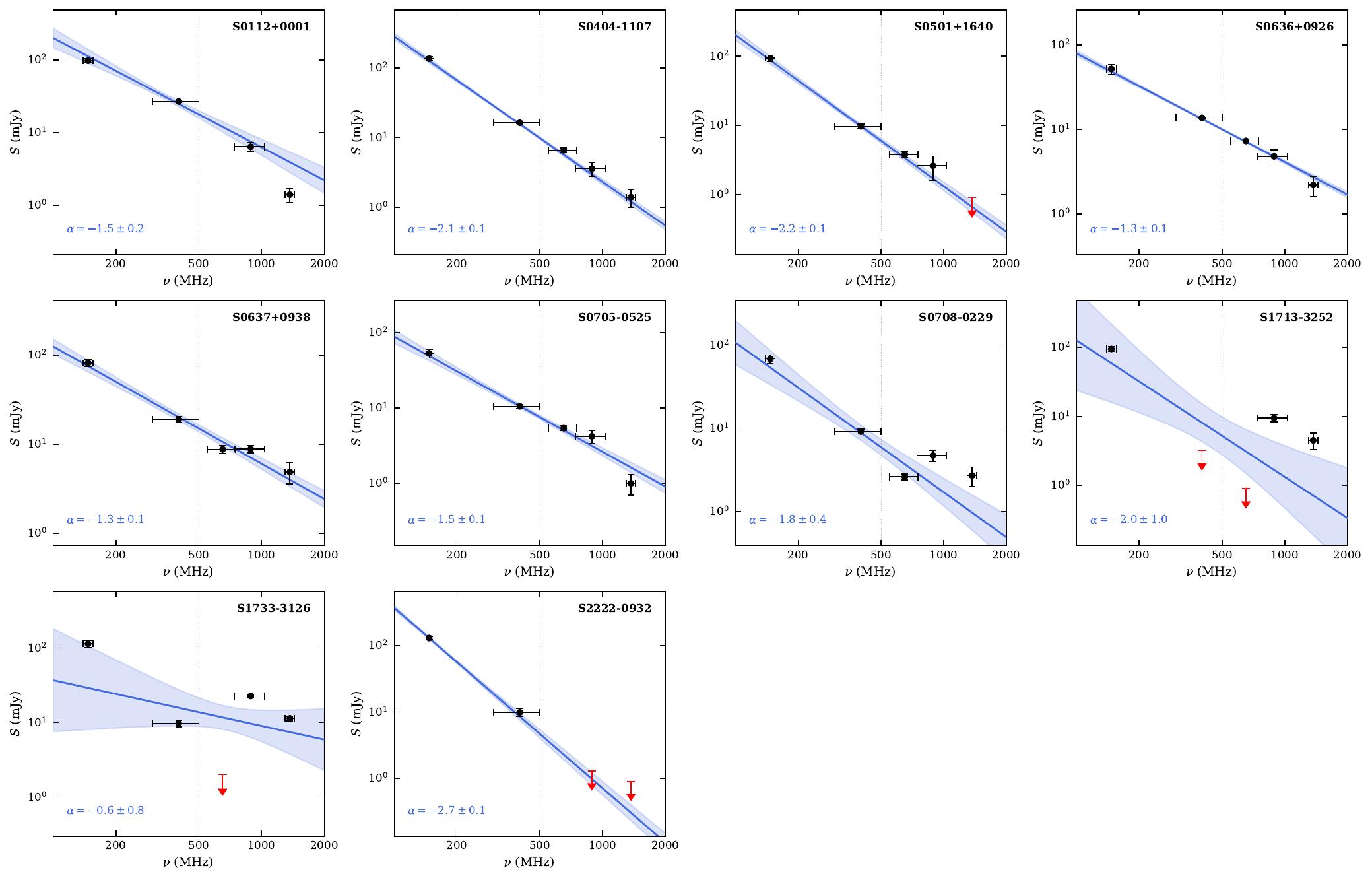}
\caption{The power-law spectral models of all the resolved sources. The black and red points represent the measured flux densities and upper limits, respectively. The best fit model is shown as a blue line, and the uncertainty region is shown by the light blue shaded region. \label{fig-spidx-resolved}}
\end{figure*}
\subsection{Spectral modelling}\label{sec-spidx}
Table~\ref{tab:flux} lists the flux density measurements and upper limits for each of the sources at 5 different frequencies, at 147\,MHz from TGSS, at 400 and 650\,MHz from our work and at 887 and 1367\,MHz from RACS. Using these measurements and upper limits, we model the radio spectrum of each source as a power law $S(\nu) = A\,(\nu/\nu_0)^{\alpha}$, with the reference frequency $\nu_0 = 500$~MHz.  As our flux density measurements as well as those from TGSS and RACS span a finite bandwidth~$\Delta\nu$, the model is band-averaged analytically over the respective bandwidths. The fitting is performed in the log space. First, the initial guesses of the two free parameters ($A$ and $\alpha$) are obtained using a weighted least square fit utilizing only the flux density measurements. Then, the best parameters are obtained by minimizing a negative log-likelihood, which is a combination of two log-likelihoods, one for the actual flux density measurements and other for the upper limits. For optimization, we use the Nelder--Mead simplex algorithm implemented in \texttt{numpy}. When the reduced chi-squared exceeds unity, we rescale the covariance by $\chi^2_\nu$ to reflect model inadequacy in the associated uncertainties. For sources with only one flux density measurement (i.e., from TGSS), the tightest spectral-index limit is derived analytically as
$\alpha \leq \ln(S_{\rm lim}/S_1) / \ln(\nu_{\rm lim}/\nu_1)$.
\par
The modeled spectra for the six compact and ten resolved sources are shown in Figures~\ref{fig-spidx-compact} and \ref{fig-spidx-resolved}, respectively. For majority of these sources, the spectra are well modelled with a single power-law. A few sources exhibit significant deviations from a single power-law, potentially due to variability or other reasons, and those are discussed in more detail in the next section. All the remaining 15 sources have flux density measurements only at one frequency (147\,MHz) and upper limits at other frequencies. The modeled spectra for these sources are shown in Figure~\ref{fig-spidx-other}.

\section{Discussion and summary} \label{sec-discussion}
The primary aim of the SCOPE survey is to find new, exciting pulsars in the sample of compact sources uncovered in radio images or those at other wavelengths, e.g., optical and X-rays, utilizing a comprehensive pulsation search approach. However, the survey also utilizes interferometric data, wherever possible, to characterize the source properties, such as the compactness and the spectrum, in the radio image domain to identify the true nature of these sources. The characterization of these properties and a comprehensive search for radio pulsations could help in uncovering interesting sources other than pulsars, e.g., HzRGs, transient or variable sources, active galactic nuclei (AGNs) with extreme properties, etc. In the following subsections, we discuss our results on an all-sky sample of 31 compact and steep spectrum sources in the context of identifying the true nature of these sources as well as implications for image-based pulsar surveys in the future.
\subsection{New pulsars}
The pulsation searches on the above sample resulted in discovery of two pulsars, \apsr and \bpsr, and the detection of another pulsar, \cpsr, which was discovered recently. All the three pulsars are MSPs, with spin periods shorter than 2.3\,ms, and their parameters are listed in Table~\ref{tab:psr_summary}. The precise locations of these pulsars listed there will help in modeling the timing behavior of these pulsars using follow-up observations. Below we briefly discuss the new discoveries.
\subsubsection{\apsr}
With a period of 1.6\,ms, \apsr is the sixth fastest rotating pulsar known to date. For the pulsar's coordinates and DM, the Galactic electron density model YMW16 \citep{YMW16} suggests the distance to be 5.7\,kpc, while the estimate from the NE2001 model \citep{CL02} is 3.7\,kpc. A more recent electron density model, NE2025 \citep{NE2025}, suggests the distance to be 5.5\,kpc, in agreement with the YMW16 model. These distance estimates put the pulsar at the edge of the Scutum-Centaurus arm of our Galaxy, though at a latitude of 7.5\mdeg. The flux density estimates obtained from our images at bands 3 and 4 of GMRT, combined with those from the TGSS and RACS images, suggest the pulsar to have a steep spectrum, with spectral index of $-2.8\pm0.5$ (see Table~\ref{tab:flux} and Figure~\ref{fig-spidx-compact}). The estimated flux density at 1.4\,GHz is only around 140\,$\mu$Jy. However, we note that the interferometric measurements from our GMRT band-4 observations at different epochs suggest that the flux density varies significantly. The flux density between two epochs separated by nearly a month varied by a factor of 2.75! A hint of the variability can also be seen in Figure~\ref{fig-spidx-compact} (see the panel marked with the source name S1840+1102) where the flux density measured at 887\,MHz (RACS-low) is higher than that measured at 650\,MHz from this work. Thus, the overall low flux density and variability (50$-$200\,$\mu$Jy at 1.4\,GHz) might have been the reasons for its non-detection in earlier pulsar surveys at L-band. We are currently following up \apsr using band-4 of GMRT and it is now established to be in a binary system with an orbital period of 1.6\,days. Further detailed study including its timing solution will be reported elsewhere \citep{Shrivastava26}.

\subsubsection{\bpsr}
The periodic radio pulsations from this source were discovered using the GBT at 820\,MHz band. Its period of 2.24\,ms firmly associate \bpsr to a gamma-ray MSP \citep{Frail18} that was earlier thought to be radio-quiet \citep{Smith23}. The three electron density models, YMW16, NE2001 and NE2025, suggest the distance to be in 3.7--4.1\,kpc range. Our flux density measurements at multiple frequencies suggest \bpsr to be extremely steep, with a spectral index of $-4.2\pm0.3$ (see Table~\ref{tab:flux} and Figure~\ref{fig-spidx-compact}). The expected flux density at 1.4\,GHz is only around 20\,$\mu$Jy, making \bpsr too weak to be detected by majority of the current radio telescopes, except for FAST\footnote{Currently, only FAST seems to have high enough sensitivity to detect this pulsar at L-band in reasonable integrations. Indeed, a few months after our discovery, an independent group detected this pulsar using FAST (priv. communication, Zhichen Pan).}. At lower frequencies, scatter-broadening comes into picture. At the center of our band-4, $\tau$ is estimated to be 2.3\,ms, i.e., more than the pulse period. At band-3, $\tau$ would be even larger. Thus, the non-detection of the periodic signal in our band-3 and band-4 observations are primarily caused by the scatter broadening being more than the pulse period. We note that our low-significant detection might have hindered identifying multiple components within the average profile, which might have resulted in overestimating $\tau$. However, the overall broadening (intrinsic as well as due to scattering) remains the likely reason for this pulsar's non-detection at lower frequencies. Having confirmed the association between the steep-spectrum source and the gamma-ray pulsar, the precise sky coordinates obtained from the radio images can now be extremely useful in obtaining the timing model using the gamma-ray data. 
\subsection{Pulsars or pulsar-like sources}
All of the sources in our sample are compact at the angular scales of the TGSS beam (25$\arcsec$). In the high angular resolution images from this work (with typical beam sizes of 8$\arcsec$ and 5$\arcsec$ at GMRT's band-3 and 4, respectively) and RACS-mid (with a median beam size of around 11$\arcsec$), 10 sources were found to be resolved. From the remaining 21 sources, 6 were found to be compact while the others could not be detected. Apart from the three pulsars that were discovered/re-discovered in our survey, another one of the source included in our sample, J1727$-$1609 is a pulsar. Its discovery is reported in the GalacticMSPs list\footnote{\url{https://pages.astro.umd.edu/~eferrara/pulsars/GalacticMSPs.txt}} and it was included in the incremental Fermi-LAT fourth source catalog \citep{aab22}. The other 17 sources remain strong pulsar candidates. While our observations have been sensitive to detect any pulsations from these sources, we note that our searches were at low radio frequencies. At these frequencies, the scatter-broadening might have been detrimental for a good fraction of the sources. All the 17 sources have spectral indices steeper than $-2.0$ (see Table~\ref{tab:flux}), except for S1835$-$1421 whose spectral index is expected to be $\leq-1.8$. From the six compact sources, S1840+1102 (or \apsr) already appears significantly variable in our band-4 images. The flux density measurements or upper limits at different frequencies indicate the sources S1727$-$1609, S1747$-$3505 and S1900+0859 also to be variable (see Table~\ref{tab:flux} and Figure~\ref{fig-spidx-compact}), and hence, strong pulsar candidates. Given the steep spectra, at higher frequencies the sources might be too faint to be detected by current radio telescopes. The 1.4\,GHz extrapolated flux densities of four compact sources, including three of the pulsars discussed above, are below 100\,$\mu$Jy. Similarly, for five of the sources that were not detected in RACS as well as in this work, the 1.4\,GHz extrapolated flux density is constrained to be around or below 100\,$\mu$Jy. Non-detection of these sources in the earlier L-band pulsar surveys can be easily explained by the lack of adequate sensitivity. While the upper limits on 1.4\,GHz flux density of the other 10 sources are around or below 1\,mJy, their actual flux densities could very well be much lower. Following these sources with sensitive telescopes like FAST or the upcoming telescope like SKA-mid, might uncover the underlying pulsars.
\subsection{Resolved sources}
Our high angular resolution images using GMRT's band-3 and band-4 revealed nine sources to be resolved. One of these nine source and one other source that could not be imaged well with GMRT, were found to be resolved in RACS images. 

The majority of these resolved sources look like double-lobed radio galaxies or diffuse emission from either a compact AGN or star-formation in a galaxy (see 
Figure~\ref{fig-resolved}). We examined the optical images of these sources using either the PanSTARRS DR1 \citep{2016arXiv161205560C} or DESI Legacy Survey DR10 \citep{2019AJ....157..168D} to look for associated host galaxies. We found that four sources (S0404$-$1107, S0637+0938, S0708$-$0229, S2222$-$0932) showed an optical galaxy somewhere midway between the peak emission from the lobes (likely to be ``hotspots'' in the FRII-type radio galaxy terminology) or associated with the diffuse, extended emission (e.g., S0112+0001). For the remaining sources, we examined their allWISE W1 and W2 infrared images \citep{2014yCat.2328....0C} and found that a likely dust-obscured host galaxy (absent in the optical images) was present for the source S0501+1640. In fact, there are likely to be two double-lobed sources in the radio image of this source in Figure~\ref{fig-resolved}, close to each other. 

Interestingly, the diffuse emission in S0636+0926 did not appear to have any associated galaxy either in the optical or in the infrared. The case of S1733$-$3126 is similarly unclear, although some emission is seen to be coincident with the radio emission in its allWISE W2 infrared image. Finally, the case of S0705$-$0525 is quite peculiar. It resembles very closely a triple radio galaxy (a lobe-core-lobe like in an FRI radio galaxy). We failed to find any optical or infrared counterpart for this source. Perhaps a more sensitive optical/infrared image is needed to identify the host galaxy in this most likely radio galaxy. Additionally, redshift and distance information for this and all extended sources are needed to confirm whether the jets or lobes extend to tens and hundreds of kpc-scales, if indeed these are radio galaxies.

Except for three sources, the spectra of these resolved sources are very well fitted with a single power law. S0708$-$0229 shows slightly higher flux densities in the RACS images than that expected from our GMRT data. The situation is much more puzzling for S1713$-$3252 and S1733$-$3126; both of these sources have flux density upper limits at intermediate frequencies that are much deeper than the spectrum implied by measurements at other frequencies. These sources are potentially resolved out at high angular resolutions of GMRT, causing the discrepancy in the flux density. The spectral indices of the well modeled sources range between $-1.3$ and $-2.7$, possibly indicating that our sample has selected the AGNs and diffuse galaxies with steeper spectra than their respective general populations.

\subsection{Implications for future image-based pulsar surveys}
The background extragalactic sources have been the major contaminants in identifying the true pulsar candidates in radio images. As discussed in the previous subsections, their spectral indices could be steep enough to confuse them with pulsars. While pulsars are truly compact sources, the extragalactic sources are resolved at reasonably high angular resolution images. In fact, target pulsar sample selected using coarse angular resolution radio images has most-likely been one of the main reasons for negligible yield in some of the large scale image-based pulsar surveys \citep[e.g.,][]{Maan18,Hyman19,Crawford21,Crawford25}. In fact, interferometric follow-up of a subset of \citet{Maan18} sample using GMRT at 1.4\,GHz has revealed that nearly 70\% of those sources are indeed resolved at 2$\arcsec$, and hence extragalactic (A. Bera et al., in prep.). In the first stage of the SCOPE survey presented here, using angular resolutions in the range 4$\arcsec$ to around 10$\arcsec$, we have identified nearly 30\% of the sources to be resolved. It can not be ruled out that a small fraction of the remaining sources might also be resolved, but interferometric follow-up of these sources with $1\arcsec - 2\arcsec$ resolutions should be able to identify nearly all the extragalactic steep spectrum sources.
\par
While high angular resolution imaging is the only definitive mean to identify and exclude the extragalactic sources from the potential pulsars, flux density might provide some help in a better selection of true pulsar candidates. The 1.4\,GHz extrapolated flux density of 60\% of all the resolved sources in our sample are above 1\,mJy, and more than 0.5\,mJy for nearly all these sources. On the other hand, the 1.4\,GHz flux density of all the compact and non-detected sources is below 1\,mJy. We note that this bimodality in the flux density is unlikely to represent a true difference in the flux density distribution of resolved and compact sources. Instead, nearly all the pulsars as bright as a mJy at 1.4\,GHz are likely to have already been discovered via blind surveys and those are excluded in the sample selection stage. Nevertheless, this apparent bi-modality could be a useful input in selecting potential pulsar candidates.
Utilizing other pulsar-like properties to identify pulsar candidates might also be useful \citep{Frail24}. One such property is the circular polarization. While a large fraction of the pulsar population is known to exhibit significant circular polarization, the mean degree of circular polarization in extragalactic sources is typically less than 1\% \citep{Macquart02}. Recently, \citet{Frail24} chose a sample of potential pulsars on the basis of circular polarization and several other pulsar-like properties. Time-domain follow-ups of this sample has already confirmed nearly 50\% of the sample to be pulsars \citep[Maan et al., in prep,][also see the SCOPE webpage\footnote{\url{http://www.ncra.tifr.res.in/~ymaan/scope.html}}]{Sengar25}! 
\par
Detection of the underlying pulsations from some of the pulsars might also be limited by the large duty cycle, either intrinsically or caused by scatter-broadening. This aspect motivates for pulsation searches at S (2--4\,GHz) and C (4--8\,GHz) bands as the scatter-broadening, and to some extent the intrinsic pulse width too, is expected to decrease significantly at higher frequencies. However, the steep spectrum sources also become very faint at such high frequencies. This poses a difficult conundrum and our discovery of \bpsr is an excellent example of this --- the pulsar is highly scatter broadened to be detected below 750\,MHz and too weak at frequencies above 1\,GHz or so. Such sources would require deep searches using sensitive telescopes like FAST or the upcoming telescopes like SKA. We also note that long integrations at higher frequencies might also not suffice to detect faint radio pulses, if, e.g., the pulsar is in a tight binary system. If the pulse broadening is not detrimental, low-frequency searches might be more suited to reveal such pulsars. Thus, in the absence of a priori information about the pulse duty cycle and scatter-broadening, follow-up at multiple frequencies in an adequate part of the spectrum remains important and almost necessary to cover the entire discovery space. Overall, designing image-based surveys using reasonably high angular resolution images and considering other pulsar-like properties, and covering adequately large parts of the radio spectrum for pulsation searches, seems to be the key to success in identifying the true pulsars. 
\par
Finally, it has also been proposed that some of the compact and steep-spectrum sources might actually be indicating towards a new Galactic source population \citep[][]{Maan18,Hyman21}. However, following the above discussion, it is of extreme importance to first probe these sources with high angular resolutions and search for pulsations across a large range of frequencies before such a conclusion can be arrived at.
\par
Summarizing, we have provided an overview of the SCOPE survey, an image-based survey for pulsars and other exotic compact objects, along with the search strategies and parameter configurations, and initial discovery from the stage-I of the survey. From an all-sky sample of 31 sources, we have reported the discoveries 2 new MSPs and detection of another MSP that was recently discovered. Using high angular resolution images obtained from simultaneously recorded interferometric data, we have also shown that nearly 35\% of the sources in our sample are resolved, hence, most-likely background extragalactic sources. The remaining sources that appear to be compact are still good pulsar candidates but a small fraction of these sources might also be HzRGs. We have also modeled the spectra of all the sources taking into account the upper limits at some of the frequencies and discussed there spectral indices in different contexts. We have also discussed our results in the context of success of future image-based pulsar surveys and strongly recommend using high angular resolution images for the sample selection and pulsation searches at a suitably large range of radio frequencies.

\begin{acknowledgments}
YM acknowledges support from the Department of Science and Technology via the Science and Engineering Research Board Startup Research Grant (SRG/2023/002657). AB acknowledges support through project CORTEX (NWA.1160.18.316) of the research programme NWA-ORC which is financed by the Dutch Research Council (NWO). PA is supported by the WISE fellowship program, which is financed by NWO. YM, DVL, YB, PK and BL acknowledge support from the Department of Atomic Energy for funding support, under project 12$-$R\&D$-$TFR$-$5.02$-$0700. 
GMRT is run by the National Centre for Radio Astrophysics of the Tata Institute of Fundamental Research. The Green Bank Observatory is a facility of the National Science Foundation operated under cooperative agreement by Associated Universities, Inc.


\end{acknowledgments}



\facilities{GMRT}
\facilities{GBT}

\software{RFIClean \citep{MvLV21}, PRESTO \citep{Ransom02}, DSPSR \citep{vSB11} }

\appendix

\section{Interferometric images and spectral modeling towards all the non-detected sources}
\begin{figure*}[ht!]
\centering
\includegraphics[scale=0.4,trim={0.0cm 0.8cm 0.2cm 0},clip]{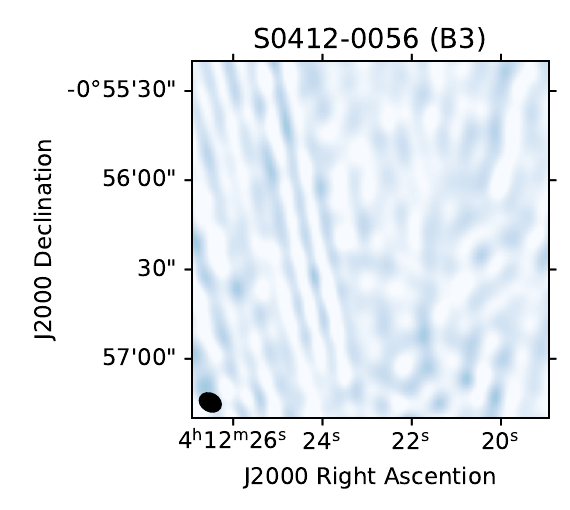}
\includegraphics[scale=0.4,trim={1.0cm 0.8cm 0.2cm 0},clip]{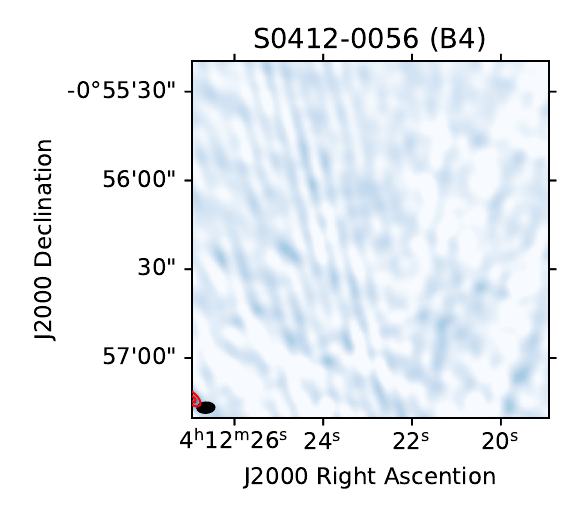}
\includegraphics[scale=0.4,trim={1.0cm 0.8cm 0.2cm 0},clip]{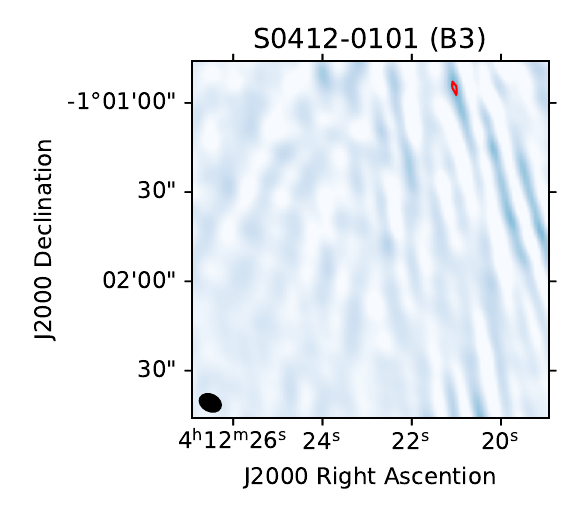}
\includegraphics[scale=0.4,trim={1.0cm 0.8cm 0.2cm 0},clip]{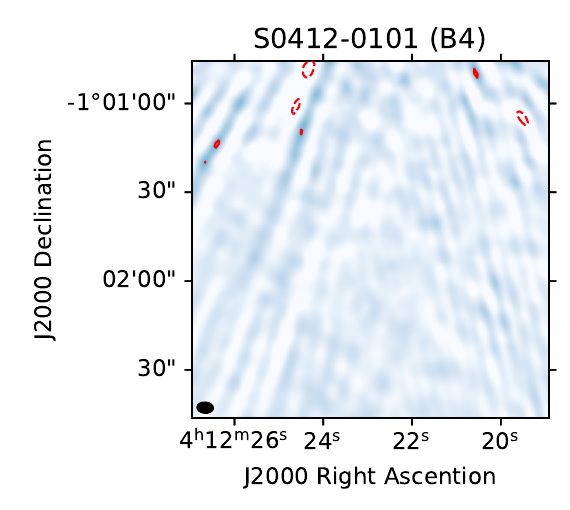}
\includegraphics[scale=0.4,trim={0.0cm 0.8cm 0.2cm 0},clip]{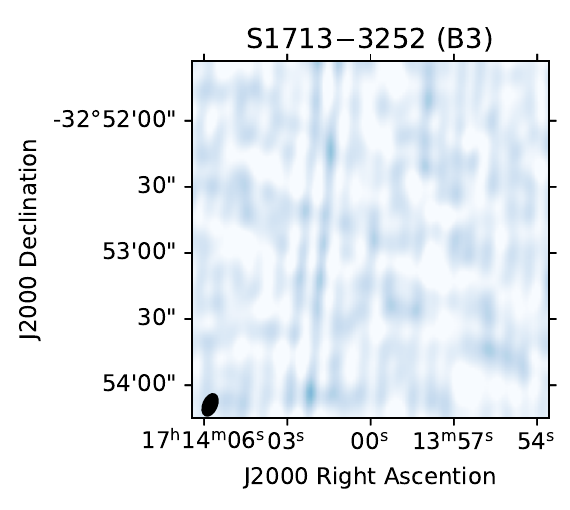}
\includegraphics[scale=0.4,trim={1.0cm 0.8cm 0.2cm 0},clip]{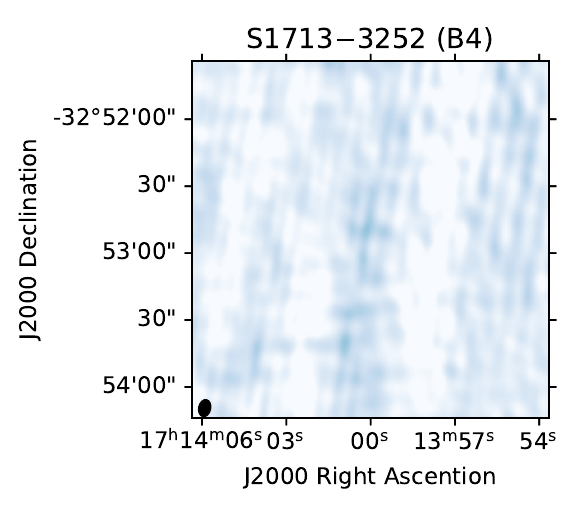}
\includegraphics[scale=0.4,trim={1.0cm 0.8cm 0.2cm 0},clip]{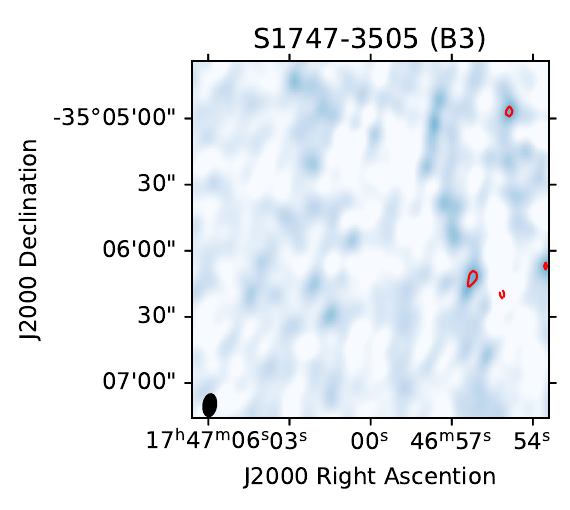}
\includegraphics[scale=0.4,trim={1.0cm 0.8cm 0.2cm 0},clip]{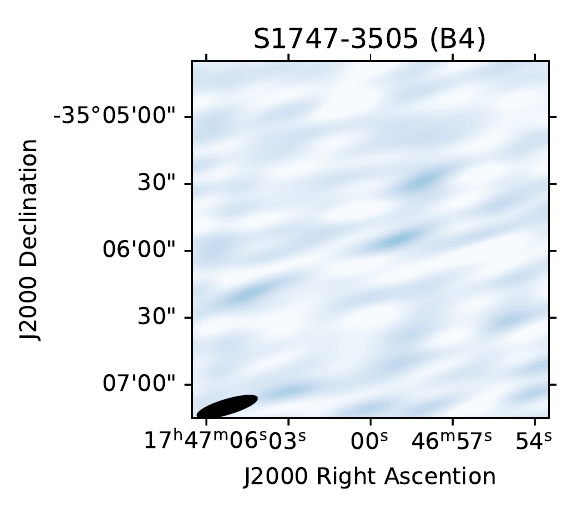}
\includegraphics[scale=0.4,trim={0.0cm 0.8cm 0.2cm 0},clip]{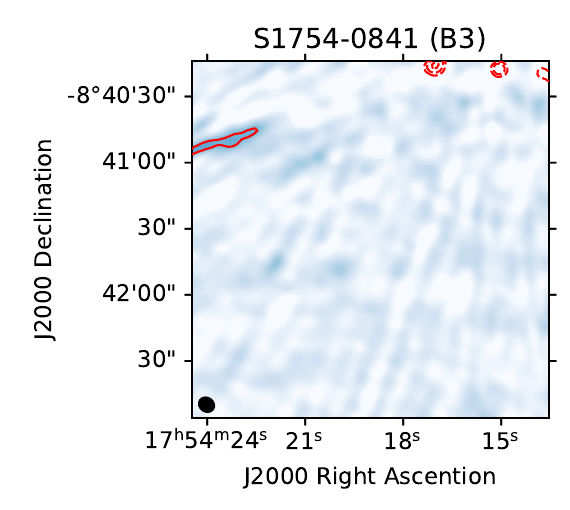}
\includegraphics[scale=0.4,trim={1.0cm 0.8cm 0.2cm 0},clip]{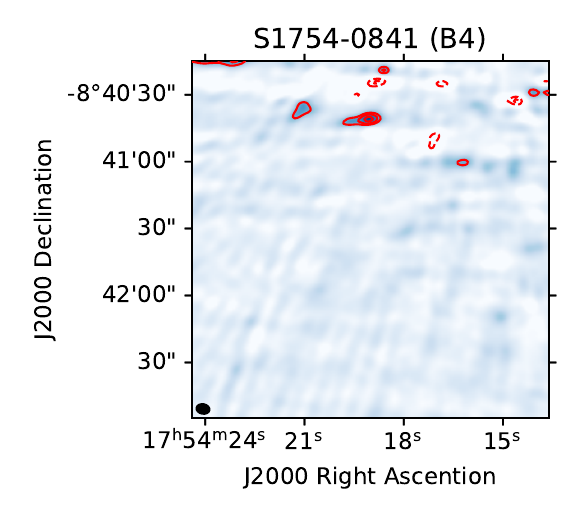}
\includegraphics[scale=0.4,trim={1.0cm 0.8cm 0.2cm 0},clip]{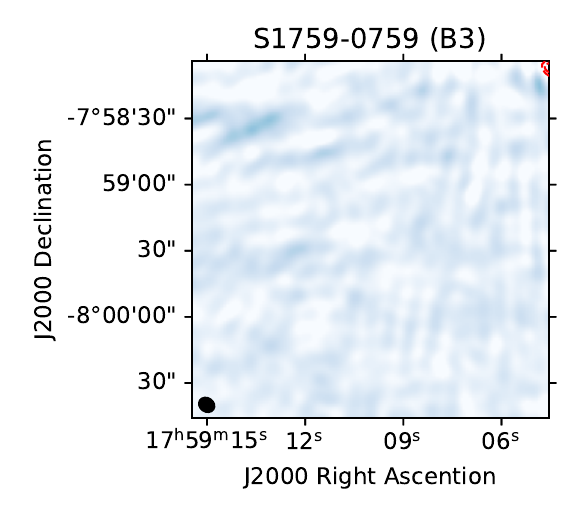}
\includegraphics[scale=0.4,trim={1.0cm 0.8cm 0.2cm 0},clip]{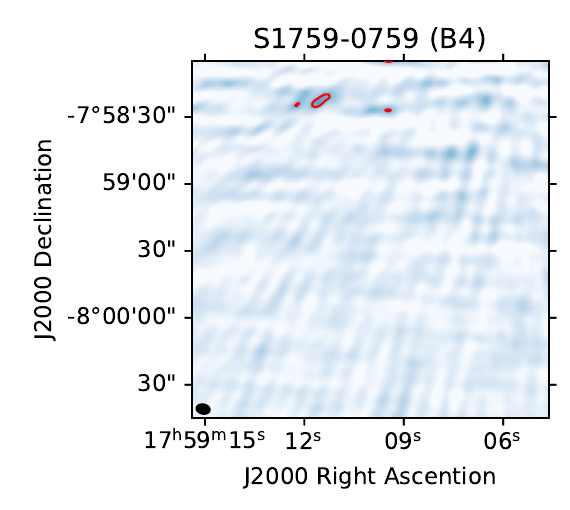}
\includegraphics[scale=0.4,trim={0.0cm 0.8cm 0.2cm 0},clip]{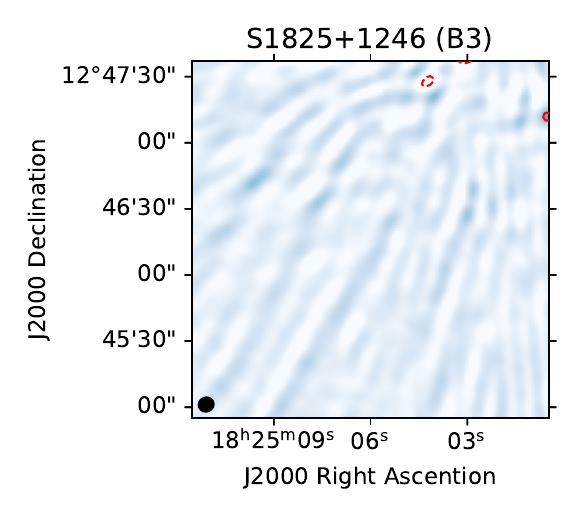}
\includegraphics[scale=0.4,trim={1.0cm 0.8cm 0.2cm 0},clip]{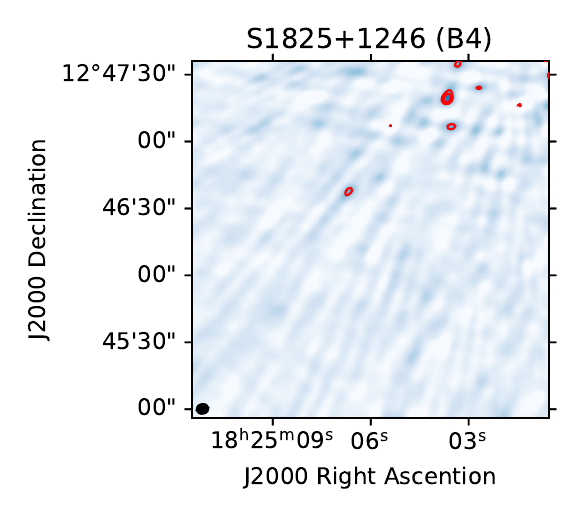}
\includegraphics[scale=0.4,trim={1.0cm 0.8cm 0.2cm 0},clip]{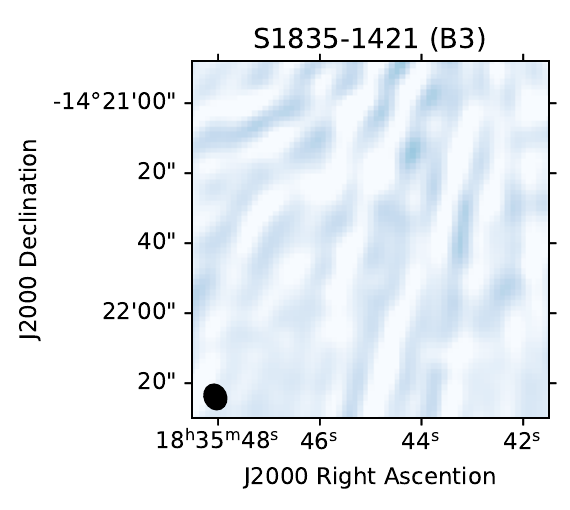}
\includegraphics[scale=0.4,trim={1.0cm 0.8cm 0.2cm 0},clip]{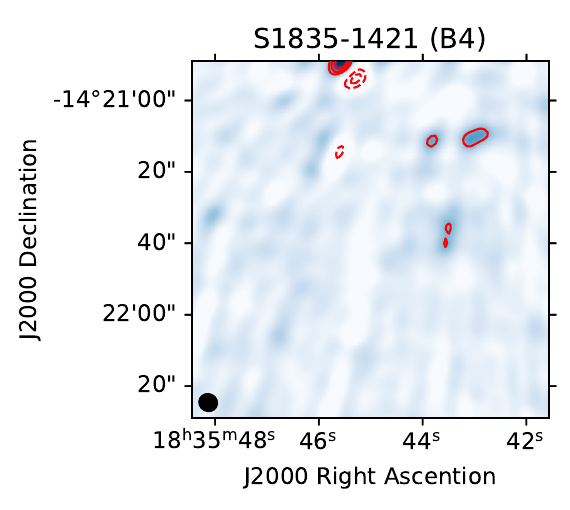}
\includegraphics[scale=0.4,trim={0.0cm 0.8cm 0.2cm 0},clip]{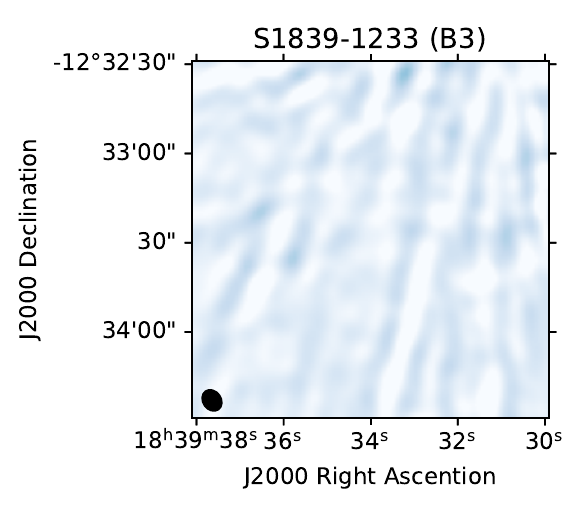}
\includegraphics[scale=0.4,trim={1.0cm 0.8cm 0.2cm 0},clip]{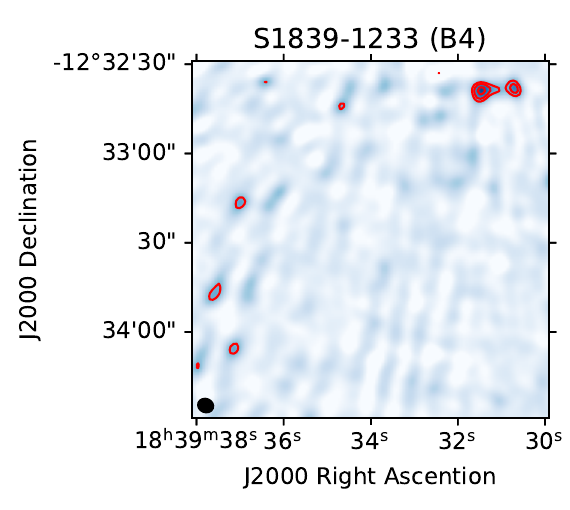}
\includegraphics[scale=0.4,trim={1.0cm 0.8cm 0.2cm 0},clip]{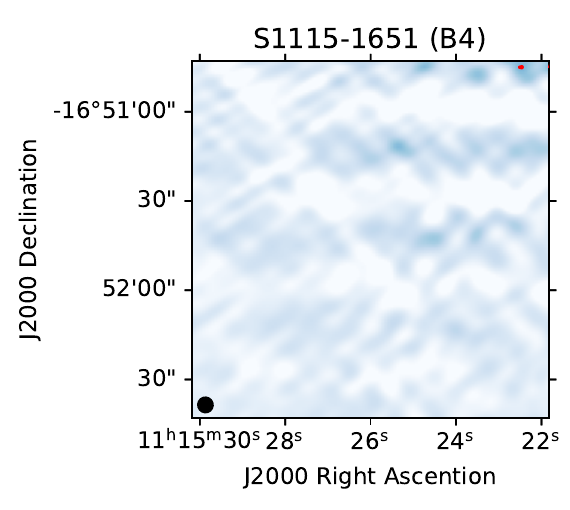}
\includegraphics[scale=0.4,trim={1.0cm 0.8cm 0.2cm 0},clip]{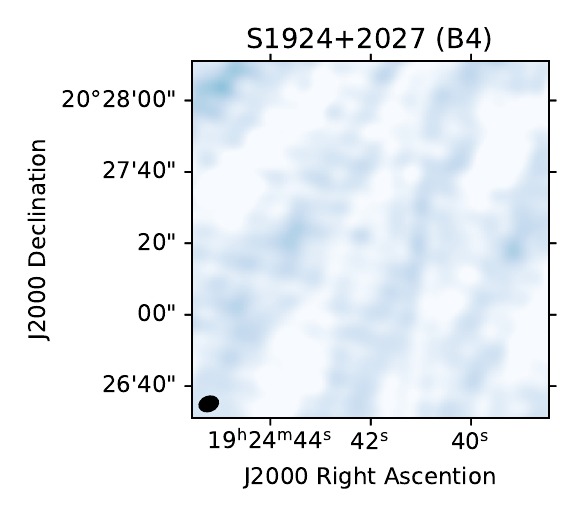}
\includegraphics[scale=0.4,trim={0.0cm 0 0.2cm 0},clip]{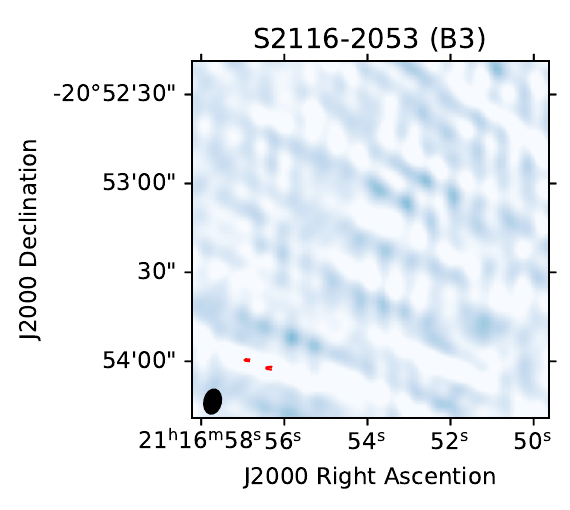}
\includegraphics[scale=0.4,trim={1.0cm 0 0.2cm 0},clip]{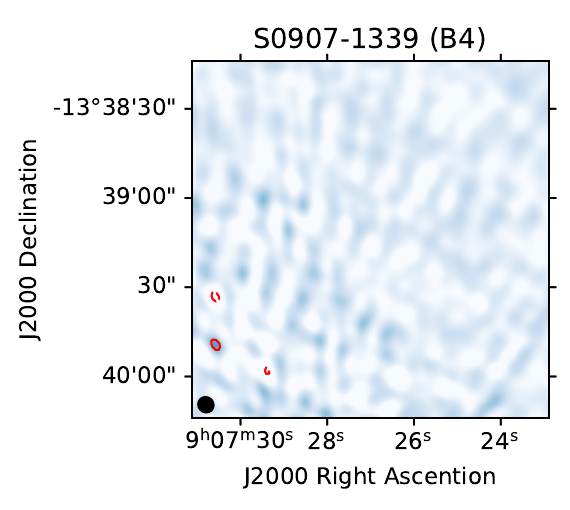}
\includegraphics[scale=0.4,trim={1.0cm 0 0.2cm 0},clip]{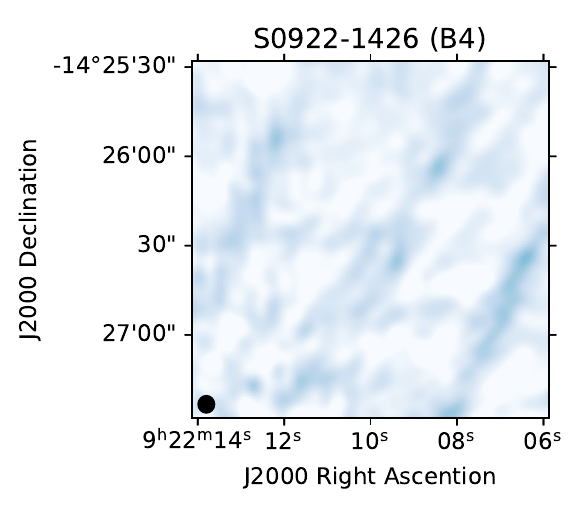}
\includegraphics[scale=0.4,trim={1.0cm 0 0.2cm 0},clip]{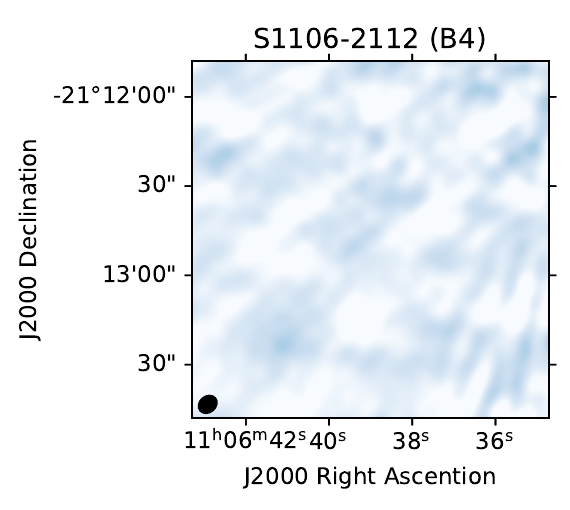}
\includegraphics[scale=0.4,trim={0.0cm 0 0.2cm 0},clip]{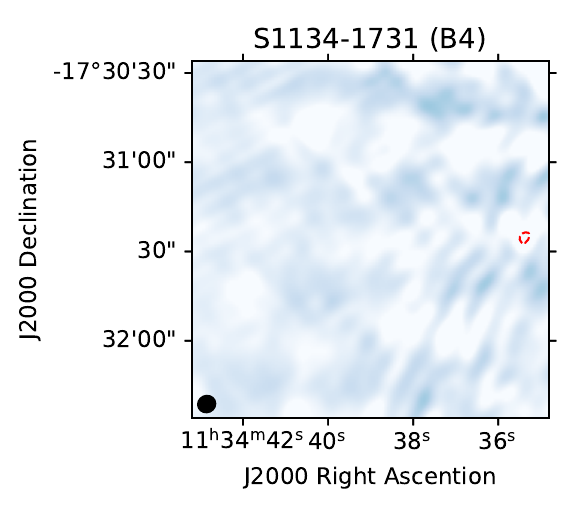}
\includegraphics[scale=0.4,trim={1.0cm 0 0.2cm 0},clip]{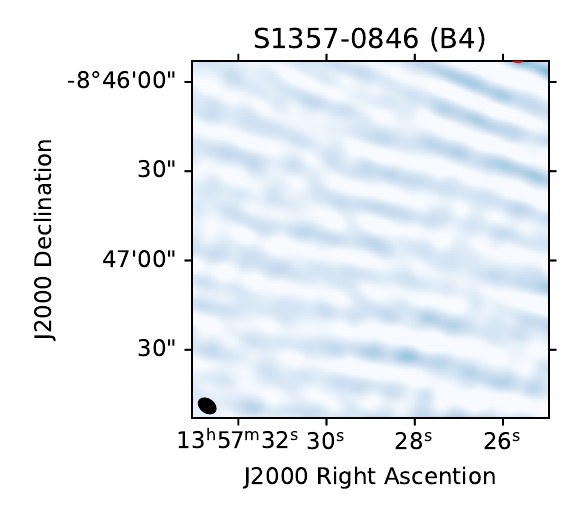}
\includegraphics[scale=0.4,trim={1.0cm 0 0.2cm 0},clip]{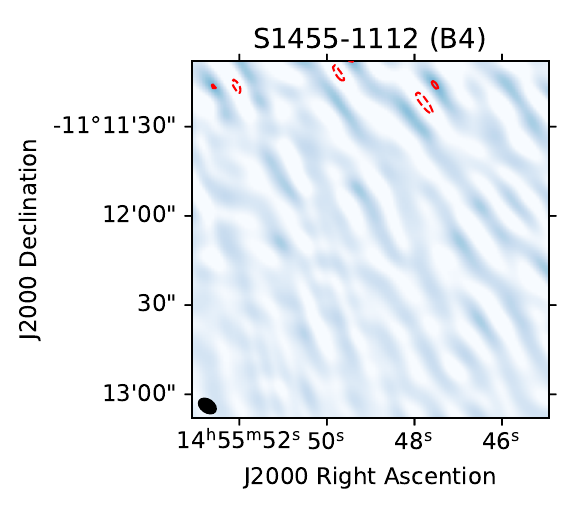}
\caption{Interferometric images of all non-detections. The color-scale is saturated at 10 times the local RMS noise. The lowest contour represents 4$\sigma_{local}$ and the contour levels increase by factors of $\sqrt{2}$.
\label{fig-nondet}}
\end{figure*}

\begin{figure*}[ht!]
\centering
\includegraphics[width=0.95\textwidth]{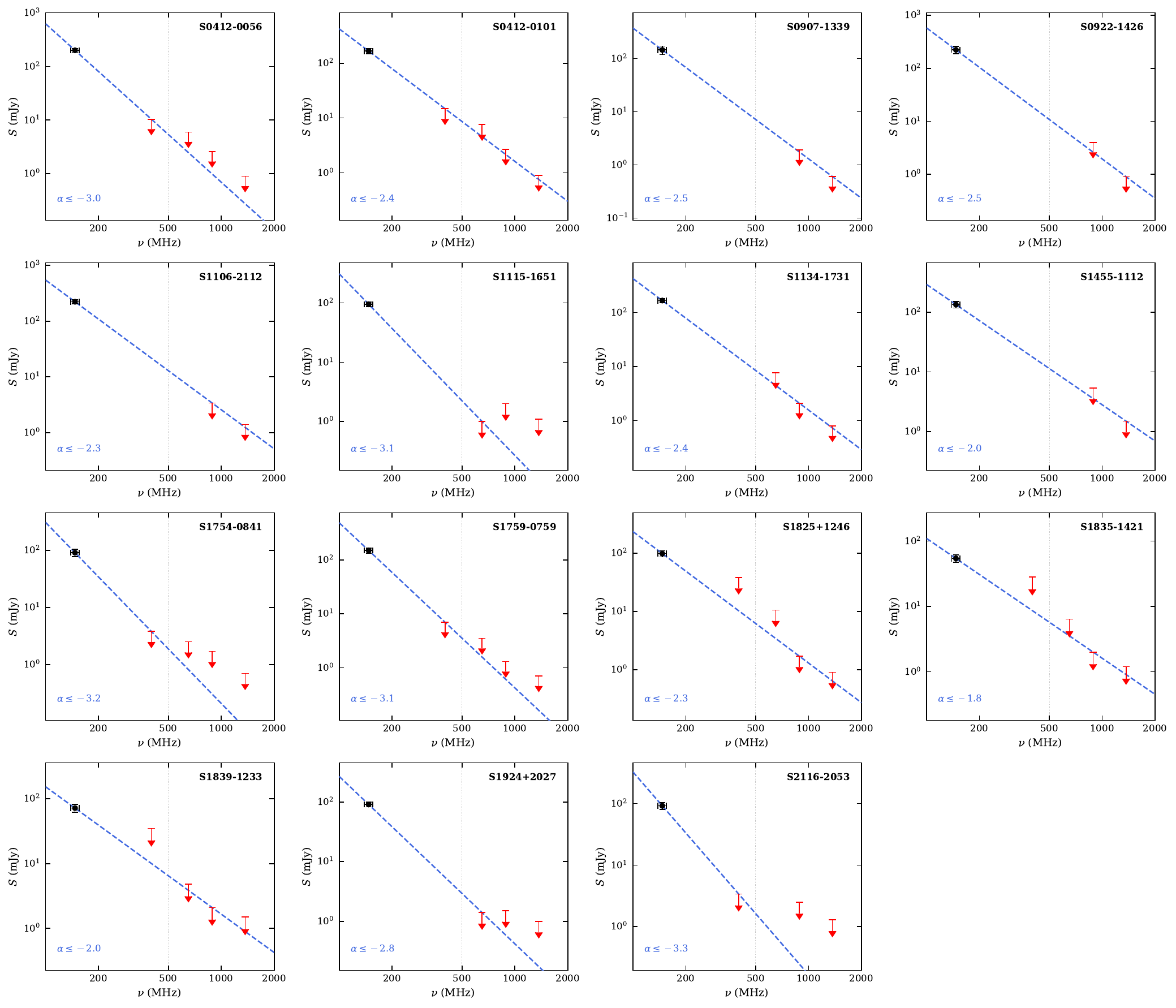}
\caption{The power-law spectral models of all the sources that are not classified as ``compact'' or ``resolved'' in Table~\ref{tab:flux}. The black and red points represent the measured flux densities and upper limits, respectively. The best fit model is shown as a blue line, and the uncertainty region is shown by the light blue shaded region. \label{fig-spidx-other}}
\end{figure*}

Figure~\ref{fig-nondet} shows the interferometric images towards all the fields where the target sources could not be detected. We note that many of the images are dominated by artifacts or deconvolution errors. However, in many cases, we are still able to obtain useful upper limits from these images. Figure~\ref{fig-spidx-other} shows the spectra modeled using the TGSS flux densities and upper limits at various frequencies for all the sources which were not detected at any of the frequencies other than 147\,MHz (TGSS).

\bibliography{references,refs}{}
\bibliographystyle{aasjournalv7}


\end{document}